%% file: paper_PRE_v2.tex
\documentclass[floatfix,aps,pra,twocolumn,superscriptaddress, longbibliography]{revtex4-2}
\usepackage{color}
\usepackage{xcolor}
\usepackage{amsmath}
\usepackage{amsthm}
\usepackage{amssymb}
\usepackage{multirow,placeins}
\usepackage[utf8]{inputenc}
\usepackage{url}
\usepackage[normalem]{ulem}

\makeatletter
\@ifundefined{textcolor}{}
{%
 \definecolor{BLACK}{gray}{0}
 \definecolor{WHITE}{gray}{1}
 \definecolor{RED}{rgb}{1,0,0}
 \definecolor{GREEN}{rgb}{0,1,0}
 \definecolor{BLUE}{rgb}{0,0,1}
 \definecolor{CYAN}{cmyk}{1,0,0,0}
 \definecolor{MAGENTA}{cmyk}{0,1,0,0}
 \definecolor{YELLOW}{cmyk}{0,0,1,0}
}

 \let\b=\beta  \let\d=\delta
   
\let\l=\lambda    
  \let\f=\varphi 
   
\let\D=\Delta   
   
   \let\io=\infty

\def\ie{{i.e. }}

\def\EE{{\cal E}} 
\def\CC{{\cal C}}\def\FF{{\cal F}} \def\HH{{\cal H}}\def\WW{{\cal W}}
 \def\BB{{\cal B}}

\providecommand{\tabularnewline}{\\}
\newcommand{\beq}{\begin{equation}} \newcommand{\eeq}{\end{equation}}

\usepackage{graphicx}
\makeatother

\begin{document}

\title{Aligning biological sequences by exploiting residue conservation and coevolution}

\author{Anna Paola Muntoni}
\affiliation{
Department of Applied Science and Technology (DISAT), Politecnico di Torino, Corso Duca degli Abruzzi 24, I-10129 Torino, Italy}
\affiliation{
Laboratoire de Physique de l'Ecole Normale Sup\'erieure, ENS, Universit\'e PSL, CNRS, Sorbonne Universit\'e, Universit\'e de Paris, F-75005 Paris, France
}
\affiliation{
Sorbonne Universit\'e, CNRS, Institut de Biologie Paris Seine, Biologie
Computationnelle et Quantitative LCQB, F-75005 Paris, France
}
\author{Andrea Pagnani}
\affiliation{
Department of Applied Science and Technology (DISAT), Politecnico di Torino, Corso Duca degli Abruzzi 24, I-10129 Torino, Italy}
\affiliation{
Italian Institute for Genomic Medicine, IRCCS Candiolo, SP-142, I-10060 Candiolo (TO) - Italy}
\affiliation{INFN, Sezione di Torino, Via Giuria 1, I-10125 Torino, Italy}

\author{Martin Weigt}
\affiliation{
Sorbonne Universit\'e, CNRS, Institut de Biologie Paris Seine, Biologie
Computationnelle et Quantitative LCQB, F-75005 Paris, France
}

\author{Francesco Zamponi}
\affiliation{
Laboratoire de Physique de l'Ecole Normale Sup\'erieure, ENS,
Universit\'e PSL, CNRS, Sorbonne Universit\'e, Universit\'e de Paris,
F-75005 Paris, France 
}

\begin{abstract}
  Sequences of nucleotides (for DNA and RNA) or aminoacids (for
  proteins) are central objects in biology. Among the most important
  computational problems is that of sequence alignment, 
  i.e.~arranging sequences from different organisms in such a way to
  identify similar regions,  to detect
  evolutionary relationships between sequences, and to predict
  biomolecular structure and function. This is typically addressed
  through profile models, which capture position-specificities like
  conservation in sequences, but assume an independent evolution of
  different positions. Over the last years, it has been well established that
  coevolution of different amino-acid positions is essential for maintaining
  three-dimensional structure and function. Modeling approaches based
  on inverse statistical physics can catch the coevolution signal in
  sequence ensembles; and they
  are now widely used in predicting protein structure, protein-protein
  interactions, and mutational landscapes. Here, we present DCAlign,
  an efficient alignment algorithm based on an approximate message-passing
  strategy, which is able to overcome the limitations of profile
  models, to include coevolution among positions in a general way,
  and to be therefore universally applicable to protein- and
  RNA-sequence alignment without the need of using complementary
  structural information. The potential of DCAlign is carefully
  explored using well-controlled simulated data, as well as real
  protein and RNA sequences.
\end{abstract}

\maketitle

\section{Introduction}

In the course of evolution, biological molecules such as proteins or RNA
undergo substantial changes in
their amino-acid or nucleotide sequences,
while keeping their three-dimensional fold
structure and their biological function remarkably conserved. In
computational biology, this structural and functional conservation is
extensively used: when we can, e.g., establish that two proteins are
homologous, i.e.~they share some common evolutionary ancestor,
properties known for one protein can be translated to its homolog
(a process known as {\it annotation}). As an example, suppose that one
is given a sequence of a human protein whose function is unknown. If
this  sequence can be properly aligned to a protein of known function
but from a different, even evolutionarily distant organism, we can
expect also the human protein to perform, globally, the same function.
Even at the finer amino-acid scale, a given position in two aligned
sequences of homologous proteins is expected to have the same
physical positioning inside the three-dimensional protein structures
of both proteins, and to share common functionality (e.g. as active
sites or in binding interfaces). 

Detecting homology is, however, not an easy task. First, homologous
proteins or RNAs may share only 20\% or even less of their residues
(i.e. amino acids or nucleotides), the
others being substituted in evolution, making the detection of
similarity rather involved. Even worse, proteins (and RNA) may change their
length, amino acids (or nucleotides) may be inserted into a sequence, or deleted from
it. Just looking to a single sequence, we have no information on which
positions might be insertions or deletions, and which positions might
be inherited from ancestors, possibly undergoing amino-acid or nucleotide
substitutions.

\begin{figure}[t]
\includegraphics[width=0.95\columnwidth]{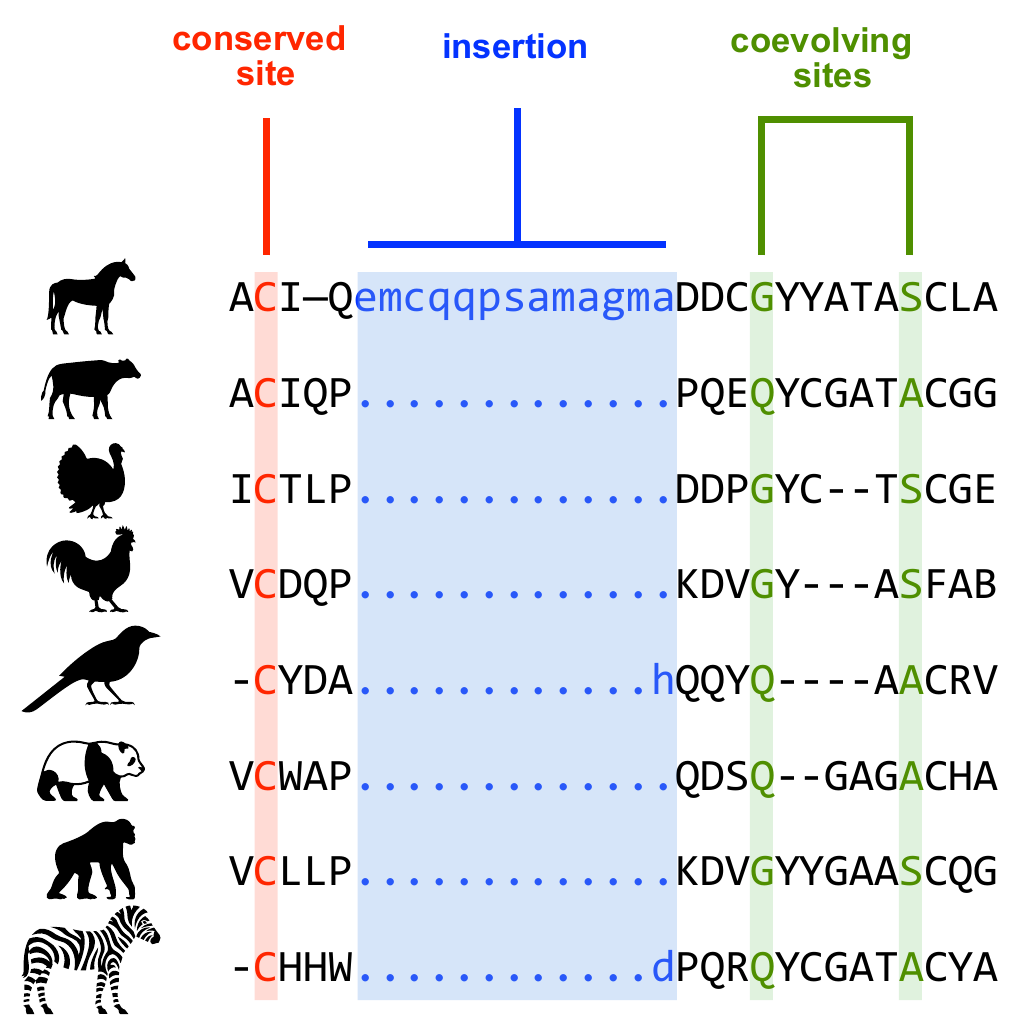}
\caption{\textbf{Example of a multiple sequence alignment.}  Sequences
  from different organisms are aligned to maximize their similarity.
  Some sites are fully conserved, others are variable. Coevolving
  sites vary in a strongly correlated way. Insertions are indicated by
  lower-case letters and dots ``.'', while deletions by dashes ``--''.
\label{fig:ex_MSA}} 
\end{figure}

To solve this problem, {\em sequence alignments} have to be
constructed \cite{durbin1998}. The objective of sequence alignments is to identify
homologous positions, also called matches, along with insertions and
deletions, such that the aligned sequences become as similar as
possible. In this context, three frequently used, but distinct
alignment problems can be identified: 
\begin{itemize}
\item {\em Pairwise alignments} compare two sequences. Under some
  simplifying assumptions, cf.~below, this problem can be solved
  efficiently using dynamic programming (i.e. an iterative method
  similar to transfer matrices or message passing in statistical
  physics)
  \cite{needleman1970general,smith1981identification}. Detecting
  homology by pairwise alignment is limited to rather close homologs. 
\item More distant homology can be detected using {\em multiple-sequence
  alignments} (MSA) of more than two sequences
\cite{edgar2006multiple}, which maximize the
  global sequence similarities by constructing a rectangular matrix
  formed by amino acids (or nucleotides) and gaps, representing both insertions and
  deletions, as illustrated in Fig.~\ref{fig:ex_MSA}. The rows of this matrix are the individual sequences, the
  columns aligned positions. Besides being able to detect more distant
  homology, MSA allow for identifying conserved positions,
  i.e. columns which do not (or rarely) change their symbol. These
  positions are typically known to be important, either for the
  functionality (e.g. active sites in proteins) or for the
  thermodynamic stability of the fold.
  Although dynamic programming methods can be generalized from pairwise to
  multiple-sequence alignments, their running time grows exponentially
  in the number of sequences to be aligned. Many heuristic strategies
  have 
  been proposed following the seminal ideas of
  \cite{feng1987progressive},
  cf.~\cite{thompson1994clustal,notredame2000t,katoh2002mafft,edgar2004muscle},
  but the 
  construction of accurate MSA of more than about $10^3$ sequences
  remains challenging.
\item For larger alignments, a simple strategy is widely used. Instead
  of constructing a globally optimal sequence alignment, {\em sequences are
  aligned one by one to a well-constructed seed MSA}
\cite{altschul1997gapped,eddy2011accelerated}. 
As in the case of MSA, this strategy allows for detecting distant homologs and for
  exploiting conservation. If the seed MSA is
  reasonably large ($\geq 10^2$ sequences) and of high quality, very
  large and rather accurate alignments of up to $10^6$ sequences can
  be constructed easily. This strategy is currently the best
  choice for constructing large families of homologous proteins
  \cite{el2019pfam} and RNAs~\cite{rfamdb}, and is also the subject of our work.
\end{itemize}
Almost all these
sequence-alignment methods are based on the simplifying assumption of
{\em independent-site evolution}. In term of sequence statistics, this amounts to
 assuming
that the global probability of observing some full-length sequence can
be factorized over site-specific but independent single-site
probabilities, also known as {\em profile models}~\cite{durbin1998}.
Profile models are thus able to capture conservation, but no coevolution. 
For an alphabet of only two symbols, hence represented by Ising spins, such a model
corresponds to a non-interacting Ising model with site-dependent local fields only; for more than
two symbols, a profile model corresponds to a non-interacting Potts model.
A successful variant are {\em profile Hidden Markov Models}
\cite{eddy1998profile}, which assume independence of matched
positions, but take into account that gap stretches are more likely
than many individual gaps to reflect the tendency of homologous
proteins (or RNAs) to accumulate, in the course of evolution, large-scale modular
gene rearrangements. 
A major advantage of profile models is their computational efficiency,
as they allow for determining optimal alignments in polynomial time
using dynamic programming (or transfer matrix) methods~\cite{durbin1998}. 
They mostly make use
of conserved positions, which serve intuitively as anchoring point in
aligning new sequences to the seed MSA. Variable positions contribute
much less to profile-based sequence alignment, and coevolution is entirely
neglected in these models.

An important exception are so-called {\em covariance models} for
functional RNA~\cite{nawrocki2013infernal}. 
Differently from proteins, RNA sequences are
characterized by low sequence conservation, 
making alignment via profile models unreliable, but highly conserved
secondary structures, due to base pairing inside the
single-stranded RNA molecules, and the formation of local helices, the
so-called {\it stems}. Base pairing does not pose constraints on the
individual bases, but on the correct pairing in Watson-Crick pairs A:U
and G:C, or wobble pairs G:U, which consequently have to be described
by a non-factorizable pair distribution. In the case of RNA, the
planar structure of the graph formed by the pairing in the RNA chains 
still enables the application of exact but
computationally efficient dynamic programming~\cite{nussinov1981,
  zuker1984, schuster1993, nawrocki2013infernal}. However, the
construction of covariance models requires the secondary structure to
be known a priori.

Suppose now that a MSA has been constructed by some alignment method.
One would then like to extract as much information as possible from this MSA.
In the last decades, amino-acid coevolution has been
established as perhaps the most important statistical 
feature of MSAs beyond conservation~\cite{de2013emerging}. 
In fact, consider a MSA of $M$ sequences $(a_1,\cdots, a_L)$, each of length $L$ (Fig.~\ref{fig:ex_MSA}).
The statistics of athe MSA is encoded into its one-site frequencies
$f_i(a)$, i.e.~the frequency of observing $a_i=a$ in a sequence,  
two-site frequencies $f_{ij}(a,b)$, i.e.~the frequency
of observing $(a_i=a, a_j=b)$ in a same sequence, three-site frequencies $f_{ijk}(a,b,c)$, and so on.
Modeling the MSA statistics by inverse
statistical physics
\cite{roudi2009,sessak2009,decelle2014,nguyen2017} 
consists in assuming that each sequence  
in the MSA is an independent realization from some unknown probability distribution
$P(a_1,\cdots,a_L)$, such that all multi-site frequencies are reproduced.
It has been realized that such a probability can be obtained by 
a maximum entropy
principle (identical to the one used in equilibrium statistical mechanics), under 
the constraint that the one-site (i.e. conservation) and two-site (i.e. coevolution) 
frequencies are correctly reproduced. The resulting $P(a_1,\cdots,a_L)$ is then a Potts model
with local fields and two-spin interactions only: interactions involving more than two spins are not present.
Remarkably, such models turn out to describe very well the three- and more-site MSA frequencies, even if these
have not been included in their construction.
This method of analysis, 
called Direct-Coupling Analysis (DCA) because it provides a set of two-spin couplings~\cite{weigt09,morcos2011,aurell2013,cocco2018},
has found
widespread applications in extracting protein-structure prediction from MSAs~\cite{marks2011, sulkowska2012,hopf2012three,ovchinnikov2017protein},
detecting protein-protein interactions
\cite{weigt09,procaccini2011dissecting, baldassi2014,feinauer16, bitbol16,gueudre16, cong2019protein, croce2019multi}, 
describing mutational effects~\cite{dwyer2013,cheng2014,figliuzzi2015,cheng2016, hopf2017mutation} or even in data-driven 
sequence optimization and design~\cite{cheng2014,reimer2019structures,russ2020evolution}. 

The resulting situation is somewhat paradoxical: MSAs are constructed using profile models (non-interacting Potts models), which 
neglect coevolution and use conservation only, but once the MSA is available, one can extract
relevant information from the coevolution signal via the DCA method, which makes use of
pairwise-interacting Potts
models. In other words, important structural and functional
information is contained in the coevolution of amino acids or nucleotides, but it is
neglected in the alignment procedure.

Our work aims at overcoming this paradox, by including the information
contained in amino-acid (or nucleotide) coevolution in aligning
sequences to a seed MSA. While this idea shows some similarity to that
of covariance models and RNA alignment, our method is much more general.
In fact, DCA modeling describes coevolution between any possible pair of
sites, and as a consequence, it allows for interactions between any possible pair of spins
in the Potts model. The resulting interaction graph is fully connected, which
makes an application of
dynamic programming (or transfer matrix) methods impossible. We cope with this problem by proposing
an approximate message passing strategy based on Belief Propagation \cite{mezard2009information},
further simplified in a high-connectivity mean-field limit for
long-range couplings~\cite{decelle2011}. We show that the resulting DCAlign algorithm
outperforms state-of-the-art alignment tools both in well-controlled simulated data and in real protein and RNA sequences.

In parallel to our work, coevolutionary models have been recently used to solve related problems, such as the remote search homology \cite{Wilburn2020.06.23.168153} and the alignment of two Potts models \cite{Talibart2020.06.12.147702}. Still, the statistical mechanics-based approach and the formalization of the problem proposed here significantly differ from those in \cite{Wilburn2020.06.23.168153,Talibart2020.06.12.147702}.

The plan of the paper is the following: we first formalize the problem
and its statistical-physics description. The latter allows us to
derive DCAlign, a combined belief-propagation / mean-field algorithm
for aligning a sequence to a Potts model constructed by DCA from the
seed MSA. The efficiency of our algorithm is first tested in the case of
artificial data, which allow us to evaluate the influence of
conservation (i.e. single-site statistics) and coevolution
(i.e. two-site couplings) in the alignment procedure. Extensive tests
and positive results are given for a number of real protein and RNA
families. Technical details of the derivation of the algorithm are
provided in the Supplementary Information (SI).

\section{Setup of the problem}

\subsection{Alignment}
\label{subsec:alignment}
The method we are going to describe can be applied to align different
types of biological sequences (viz.  DNA, RNA, proteins). We discuss here
the protein case, but the extension to the other cases is
straightforward and it will be considered below. Let us consider an amino-acid sequence
$\boldsymbol{A}=\left(A_{1},\ldots,A_{N}\right)$ containing a protein
domain $\boldsymbol{S} = (S_1,...,S_L)$ of a known family, which we
want to identify.  Note that $\boldsymbol{S}$ may contain
amino acids and gaps, while the original sequence $\boldsymbol{A}$ is
composed exclusively by amino acids.
We assume that the protein family is well described
by a Direct Coupling Analysis (DCA) model, or Potts Hamiltonian, or
simply ``energy'',
\begin{equation}
{\cal H}_{\mathrm{DCA}}\left(\boldsymbol{S}\, | \,
\boldsymbol{J},\boldsymbol{h}\right)=-\sum_{ i<j }^{1,L}J_{ij}
\left(S_{i},S_{j}\right)-\sum_{i=1}^Lh_{i}\left(S_{i}\right)\ . \label{eq:potts}
\end{equation}
Here, the sequence $\boldsymbol{S} = (S_1,...,S_L)$ is assumed to be
aligned to the MSA of length $L$ of the protein family, and the set of
parameters $\boldsymbol{J}$ and $\boldsymbol{h}$ are considered as
known, having been learned from some seed alignment~\cite{figliuzzi2018pairwise}. The energy ${\cal
  H}_{\mathrm{DCA}}$ is then considered as a ``score'' (lower energy
corresponds to higher score) for sequence $\boldsymbol{S}$ to belong
to the protein family.  We address the problem of
aligning a sequence $\boldsymbol{A}$ to the model ${\cal
  H}_{\mathrm{DCA}}$ or, in other words, of detecting the domain in
$\boldsymbol{A}$ that has the best score within the model ${\cal
  H}_{\mathrm{DCA}}$. In this setting, the solution to our problem is
the sub-sequence (cf. below for the precise definition of a
sub-sequence including insertions and deletions) that, among all the
possible sub-sequences of $\boldsymbol{A}$, minimizes the energy
(or, at finite temperature, is a typical sequence sampled from ${\cal H}_{\mathrm{DCA}}$).  
The energy
thus serves as a cost function for comparing different candidate
alignments.  Contrarily to profile models, which only take into
account conservation of single residues, DCA models also include
pairwise interactions related to residue coevolution (and thus in
particular at any linear separation along the sequence
$\boldsymbol{A}$), hopefully leading to more accurate alignments in
cases where conservation alone is insufficient.

However, the DCA model $\mathcal{H}_{\mathrm{DCA}}$ does not model
insertions, because the parameters $\boldsymbol{J}$ and
$\boldsymbol{h}$ are inferred from a seed MSA where all columns
containing inserts have been removed. A suitable additional cost has
thus to be assigned to amino-acid insertions, which are needed in
order to find a low-cost alignment. Still, we have to prevent our
algorithm from picking up energetically favorable but isolated amino acids
out of the (possibly long) input sequence $\boldsymbol{A}$. For
modeling this cost, we will explicitly refer to the insertion
statistics in the full seed alignment.

Note that the DCA model contains position-specific gap terms in the
$\boldsymbol{J}$ and $\boldsymbol{h}$; so the gap statistics of the
seed MSA is fully described by the DCA model alone. Nonetheless, the
observed statistics deeply depends on how the seed is constructed, and
it could a priori be non-representative of the gap statistics of the
full alignment.  To take into account this degree of variability, we
allow for the introduction of an additional energy term associated
with the presence of gaps. A more detailed discussion is reported in
Sec.~\ref{sec:penalties}.

Formally speaking, the alignment problem reduces to finding a sub-sequence
$\boldsymbol{S}=\left(S_{1},\ldots,S_{L}\right)$ 
of $\boldsymbol{A}=\left(A_{1},\ldots,A_{N}\right)$ such that:
\begin{enumerate}
\item the sub-sequence $\boldsymbol{S}$ forms an ordered list of amino-acids
in $\boldsymbol{A}$ (``match'' states) with the possibility of
(a) adding gaps, denoted as ``--'', between two consecutive positions
in $\boldsymbol{A}$ and (b) skipping some amino acids of $\boldsymbol{A}$,
\ie interpreting them as insertions; 
\item the aligned sequence $\boldsymbol{S}$ minimizes
\begin{equation}\label{eq:Edef}
\begin{split}
 E(\boldsymbol{S} | \boldsymbol{J}, & \boldsymbol{h}, \boldsymbol{\lambda}, \boldsymbol{\mu}) = 
{\cal H}_{\mathrm{DCA}}\left(\boldsymbol{S}\, | \, \boldsymbol{J},\boldsymbol{h}\right) \\ 
&+ \HH_{\rm{gap}}(\boldsymbol{\mu}) + \HH_{\rm{ins}} (\boldsymbol{\lambda}) \ ,
\end{split}
\end{equation}
being $\HH_{\rm{gap}}(\boldsymbol{\mu}) $ a penalty for adding gaps
that depends on the number and position of gaps and on the
hyper-parameters $\boldsymbol{\mu}$, and $\HH_{\rm{ins}}
(\boldsymbol{\lambda})$ being a penalty on insertions, parametrized by
the hyper-parameters $\boldsymbol{\lambda}$.
\end{enumerate}
We first consider here the case where $N \gtrsim L$, \ie we are trying
to align a domain in a longer sequence, or $N < L$ when we are trying
to align a fragment.  The case $N\gg L$, \ie when we search for a hit
of the DCA model in a long sequence, may be computationally hard for
the approach proposed here, because the alignment time scales roughly
as $L^2 N^2$, as discussed below.  Current state-of-the-art alignment
methods like BLAST use heuristics to approximately locate possible
hits, and perform accurate alignment search only in these restricted
regions, to speed up search. It might be necessary to do this before
running DCAlign, but this is not the objective of the current work. In
general, it will be better if $N$ is not too different from $L$.

\begin{figure}[t]
\includegraphics[width=0.95\columnwidth]{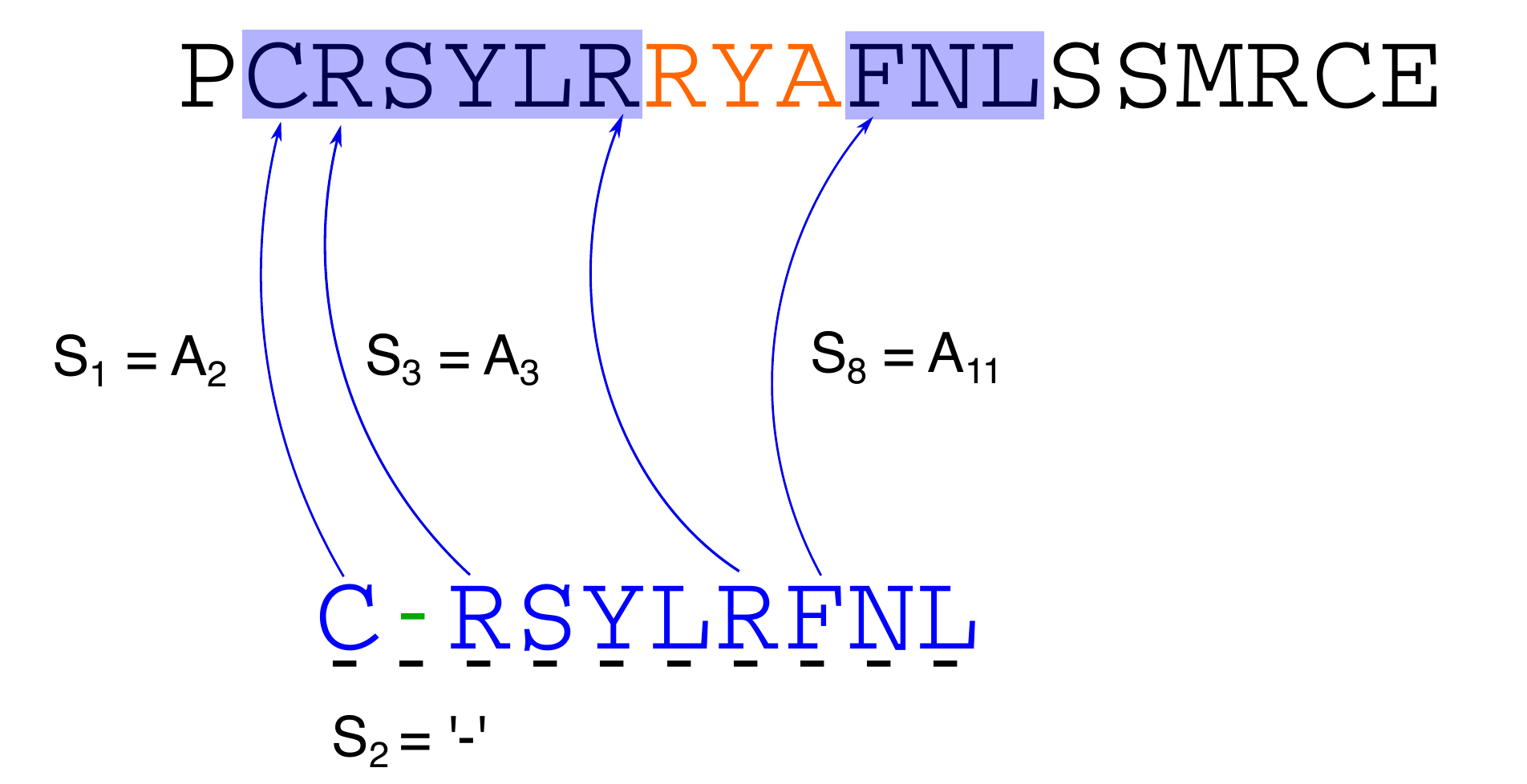}
\caption{\textbf{Example of an alignment.} On the top, we show a full
  length sequence $\boldsymbol{A}$, and on the bottom its alignment
  $\boldsymbol{S}$, in which both gap and insertion events occur. The
  domain to be aligned is highlighted in blue. We show explicitly
  three matched states $S_{1} = A_{2}$, $S_{3} = A_{3}$ and $S_{8} =
  A_{11}$ and a gap insertion in $S_{2} =$``--". In this example we
  also show a possible way of skipping some amino-acids in the
  original sequence, that is to assign three insertions, highlighted
  in red. \label{fig:ex_align}} 
\end{figure}

An example of a full sequence and its alignment is given in Fig.~\ref{fig:ex_align}. The sequence on the top is a full sequence of $N = 19$ amino-acids, and the highlighted part is the target domain. The aligned sequence of length $L = 10$, reported on the bottom, consists in one gap in position $2$ and $9$ matched amino-acids. 
 

\subsection{Gap and insertion penalties}
\label{sec:penalties}

The hyper-parameters $\boldsymbol{\mu}$, $\boldsymbol{\lambda}$
determine the cost of adding a gap or an insertion in the aligned
sequence. They must be carefully determined to allow for these events
without affecting the quality of the alignment. In other words, we
would like to reduce, as much as possible, the number of gaps (when
the statistics of gaps of the seed we use is biased) and to
parsimoniously add insertions when energetically favorable, avoiding
to pick up isolated amino acids in the alignment.

To deal with insertions we use a so-called affine penalty
function \cite{durbin1998} parametrized by $\lambda_{i}^{o}$, the cost
of adding a first insertion between positions $i-1$ and $i$,
and $\lambda_{i}^{e}$, the cost of extending an existing insertion,
with $\lambda_{i}^{o} > \lambda_{i}^{e}$. This results in
\beq\label{eq:HHins}  
\begin{split}
\HH_{\rm{ins}} (\boldsymbol{\lambda}) &= \sum_{i=2}^L \f_i\left(\Delta n_{i}\right) \ , \\
\f_i\left(\Delta n_{i}\right)&= \left(1-\delta_{\Delta n_{i},0}\right)
\left[\lambda_{o}^{i}+\lambda_{e}^{i}\left(\Delta n_{i}-1\right)\right] \ ,
\end{split}\eeq
where $\Delta n_i$ is the number of insertions between positions $i-1$
and $i$.  This set of parameters can be learned from a seed alignment
through a Maximum Likelihood (ML) approach as reported in
Sec.~\ref{sec:MLinspen}.  Finally, we introduce two types of gap
penalties, denoted by $\mu^{\rm{int}}$ and $\mu^{\rm{ext}}$, which are
associated with an ``internal" gap between two matched states and with
an ``external" gap (at the beginning and at the end of the aligned
sequence), respectively.  This gives \beq
\HH_{\rm{gap}}(\boldsymbol{\mu}) = \sum_{i=1}^L \mu_i \ , \eeq where
$\mu_i=0$ for match states, $\mu_i = \mu^{\rm{int}}$ for internal gaps
and $\mu_i = \mu^{\rm{ext}}$ for external gaps.

An illustration is given by the aligned sequence in
Fig.~\ref{fig:ex_align}. Insertions are highlighted in red, and the
total insertions penalty is then given by
$\lambda_{o}^{8}+\lambda_{e}^{8}+\lambda_{e}^{8}$.  A gap, which
increases the total energy by $\mu^{\rm{int}}$, is highlighted in
green at position 2 of $\boldsymbol{S}$.

\subsection{Statistical physics model}

We now want to construct a discrete statistical-physics model which
defines this alignment.  For the positions ${1\leq i\leq L}$, the
model has to encode the position of the gaps and of the match states,
with their corresponding symbol in the sequence $(A_{1},...,A_{N})$.
We therefore introduce two variables per site $1\leq i\leq L$. The
first one is a boolean ``spin'' $x_{i}\in\{0,1\}$, which tells us if
$S_{i}$ is a gap ($x_{i}=0$) or an amino-acid match ($x_{i}=1$). The
second one is a ``pointer'' $n_{i}\in\{0,...,N+1\}$, which gives, for
the case of match states $x_{i}=1$ and $1 \leq n_{i} \leq N$, the
corresponding position in the original sequence
$(A_{1},...,A_{N})$; note that this allows for insertions if
$n_{i+1}-n_i>1$. If matched symbols start to appear only from a
position $i > 1$, we then fill the previous positions $\{j: 1\leq
j<i\}$ with gaps having pointer $n_{j} = 0$. Similarly, if the last
matched state appears in $i < L$ we fill a stretch of gaps in
positions $\{j:i< j\leq L\}$ having $n_{j} = N+1$.  This encoding
allows one to distinguish the ``external'' gaps at the boundary of the
aligned sequence, whose total number we denote as
$N_{\rm{gap}}^{\rm{ext}}$, from the ``internal" ones, \ie between two
consecutive matched states, whose total number is
$N_{\rm{gap}}^{\rm{int}}$.  Formally, the number of gaps and insertions are
\begin{eqnarray}
N_{\rm{ins}} &=& \sum_{i=1}^{L-1} (n_{i+1} - n_i -1) \mathbb{I}[N+1>n_{i+1}  > n_i  >0]   \ , \nonumber \\
\label{eq:gaps} 
N_{\rm{gap}}^{\rm{int}}&=&\sum_{i=1}^L\delta_{x_{i},0}\mathbb{I}[0<n_{i}<N+1] \ , \\
N_{\rm{gap}}^{\rm{ext}}&=&\sum_{i=1}^L\delta_{x_{i},0}(\delta_{n_{i},0} + \delta_{n_{i},N+1}) \ ,  \nonumber
\end{eqnarray}
where $\mathbb{I}[\EE]$ is the indicator function of the event $\EE$.
Introducing the short-hand notation $A_{0}=$ ``--'' (gap), a model
configuration $(x_{1},...,x_{L},n_{1},...,n_{L})$ results in an
aligned sequence $(S_{1},...,S_{L})=(A_{x_{1}\cdot
  n_{1}},...,A_{x_{L}\cdot n_{L}})$.  The auxiliary variables
$(\boldsymbol{x}, \boldsymbol{n})$ must be additionally assigned such
that the positional constraints illustrated in Fig.~\ref{fig:ex_align}
are satisfied, \ie the target sub-sequence must be ordered, as we now
describe.

First of all, in order
to correctly set the pointers in presence of gaps in the first and last positions, it is sufficient to set the state of node $i = 1$ as 
\begin{eqnarray}
n_{1} = 0 & \ {\rm if}\  & x_{1}=0 \ , \nonumber \\
N+1 > n_{1} > 0 &\  {\rm if}\  & x_{1}=1 \ , 
\end{eqnarray}
and the state of node $i = L$ as
\begin{eqnarray}
n_{L} = N+1 &\  {\rm if}\ & x_{L}=0 \ , \nonumber \\
N+1 > n_{L} > 0 &\ {\rm if}\ & x_{L}=1\ .
\end{eqnarray} 
 These properties can be formally expressed by the following two single-position constraints 
\beq\begin{split}
  \label{boundconstr}
\chi_{\mathrm{in}}(x_1,n_1)  & =  \d_{x_1,0} \d_{n_1, 0} \\ &+ \d_{x_1,1} (1-\d_{n_1,0})(1-\d_{n_1,N+1})   \ , \\
\chi_{\mathrm{end}}(x_L,n_L) & =  \d_{x_L,0} \d_{n_L, N+1} \\ &+ \d_{x_L,1} (1-\d_{n_L,0}) (1-\d_{n_L,N+1})  \ .
\end{split}\eeq

\begin{figure}[t]
	\includegraphics[width=0.95\columnwidth]{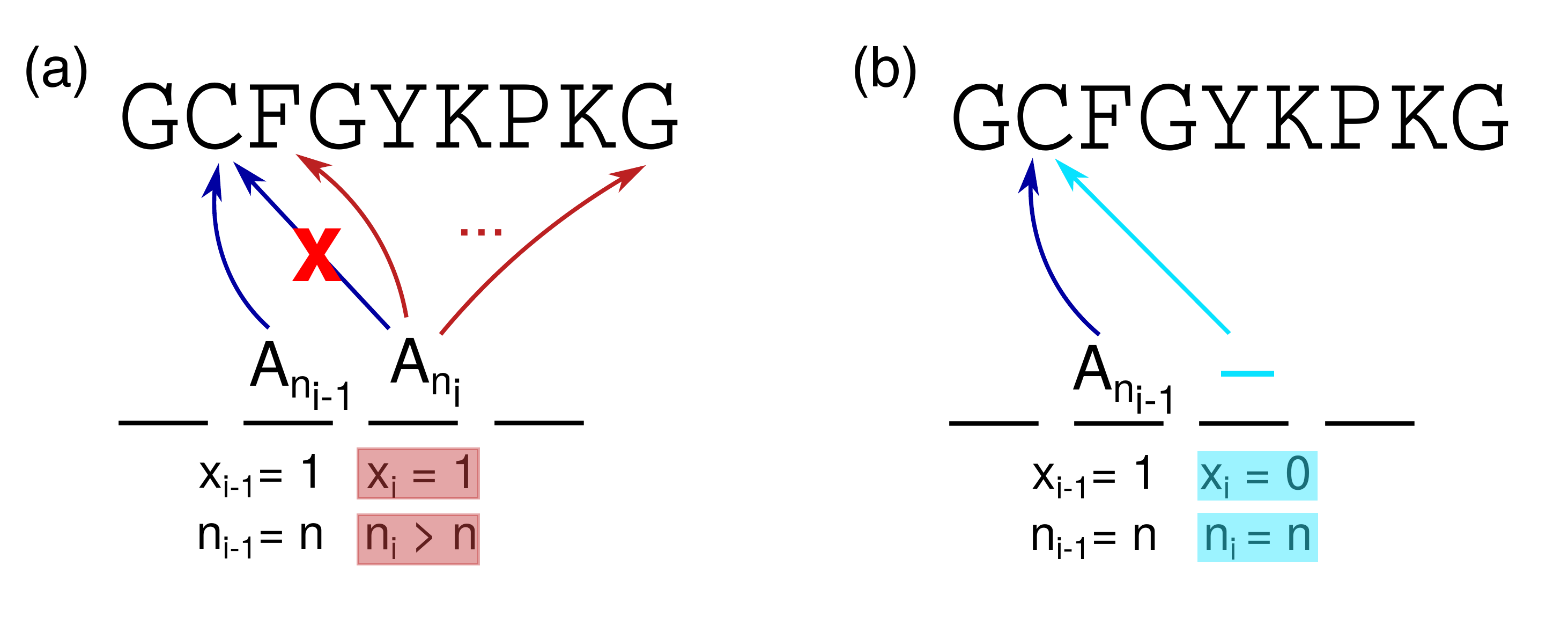}
	\caption{\textbf{Short range constraints.} We plot in (a) a
          feasible assignment of two consecutive matched states. If in
          position $i-1$ we assign $S_{i-1} = A_{n}$, we can then
          align the next position $i$ to one of the possible
          amino-acids $S_{i}\in\{A_{n+1},...,A_{N}\}$. As a
          consequence $\left(x_{i-1}, x_{i}\right) = \left(1,
          1\right)$, $n_{i-1} = n$ while $n_{i}\in\{n+1,...,N\}$. In
          (b) we plot a feasible inclusion of a gap in position
          $i$. If the previous site $i-1$ points to $n_{i-1} = n$, we
          then assign $\left(x_{i}, n_{i}\right) = \left(0,n\right)$
          to keep memory of the aligned sequence. In the next
          position, $i+1$, we can match an amino-acid in further
          positions (according to the constraint in (a)) or add
          another gap with pointer $n_{i+1} =
          n$. \label{fig:contraints}}
\end{figure}

Next, we need to locally impose that, for each position  $1<i<L$,
\begin{eqnarray}
n_{i}=n_{i-1} &\ {\rm if}\ & x_{i}=0 \text{ and } n_i < N+1 \ , \nonumber \\
n_{i}>n_{i-1} &\ {\rm if}\ & x_{i}=1 \text{ or } n_i = N+1\ , 
\end{eqnarray}
\ie the pointer $n_{i}$ remains constant when $x_{i}=0$,
and it jumps to any later position in $n_{i-1}+1,...,N$ if $x_{i}=1$. 
This jump, besides determining the amino-acid $S_{i}$ to be placed in
position $i$, also allows for identifying inserts according to
Eq.~\eqref{eq:gaps}. A pictorial representation of this constraint is
shown in Fig.~\ref{fig:contraints}. 
We can formally encode these constraints in a ``short-range" function
$\chi_{\rm{sr}}(x_{i-1}, n_{i-1}, x_{i}, n_{i})$  
that, for each pair of consecutive positions $(i-1,i)$, indicates the
feasible/unfeasible configurations of the variables $(x_{i-1},
n_{i-1}, x_{i}, n_{i})$ 
and the associated cost of insertions, as
\beq\label{eq:cond}\begin{split}
\chi_{\mathrm{sr}}(0,n_{i-1},0,n_i) & =  \mathbb{I}(n_{i}=n_{i-1}) \ ,    
\\ 
\chi_{\mathrm{sr}}(1,n_{i-1},0,n_i) & = \mathbb{I}(n_{i}=n_{i-1}\vee n_{i}=N+1) \ ,
\\
\chi_{\mathrm{sr}}(0,n_{i-1},1,n_i) & = e^{-\f_{i}\left(\Delta n_i\right) \mathbb{I}(n_{i-1}>0)}\times \\&\times \mathbb{I}(0 \leq n_{i-1}<n_i<N+1) \ ,
\\
\chi_{\mathrm{sr}}(1,n_{i-1},1,n_i) & =e^{-\f_{i}\left(\Delta n_i\right)}\times \\&\times \mathbb{I}(0 < n_{i-1}<n_i<N+1) \ ,
\end{split}\eeq
where  the function $\f_i(\Delta n_i)$ is the contribution of the ${i\rm{-th}}$ position to the affine insertion penalty, as given in Eq.~\eqref{eq:HHins} with
$\Delta n_{i} = n_{i} - n_{i-1} -1$.
Note that combining 
the constraints in Eq.~\eqref{boundconstr} and Eq.~\eqref{eq:cond}, positions $\{j > 1\}$ ($\{j < L\}$) can either have a gap with $n_{j} = 0$ (${n_{j} = N+1}$) or the first (last) match at any position $n_{j} > 0$ ($n_{j} < N+1$) with no insert penalty. 

Finally,
gap penalties can be encoded in a single-variable weight,
\begin{equation}
\chi_{\mathrm{gap}}\left(x_{i},n_{i}\right)	=	e^{-\left(1-x_{i}\right)\mu\left(n_{i}\right)} \ ,
\end{equation}
with
\begin{equation}
\mu\left(n\right)=\begin{cases}
\mu^{\mathrm{ext}} & n=0 \vee n=N+1 \ , \\
\mu^{\mathrm{int}} & 1\leq n \leq N \ .
\end{cases}
\end{equation}

The DCA Hamiltonian can be rewritten in terms of the auxiliary variables as
\beq\begin{split}
&\HH_{\rm DCA}(\boldsymbol{x}, \boldsymbol{n} | \boldsymbol{J}, \boldsymbol{h}) =  \\ &\ \ 
-  \sum_{i<j}J_{ij}(A_{x_{i}\cdot n_{i}},A_{x_{j}\cdot n_{j}}) - \sum_{i}h_{i}(A_{x_{i}\cdot n_{i}})  \ ,
\end{split} \eeq
while the
global cost function $E$ in Eq.~\eqref{eq:Edef} is
\beq\begin{split}
E(\boldsymbol{x},& \boldsymbol{n} | \boldsymbol{J}, \boldsymbol{h}, \boldsymbol{\lambda}, \boldsymbol{\mu})  = \\
&\HH_{\rm DCA}(\boldsymbol{x}, \boldsymbol{n} | \boldsymbol{J}, \boldsymbol{h}) 
   + \sum_{i} (1-x_{i})\mu(n_{i}) \\ 
   &+ \sum_{i=2}^L  \f_{i}\left(\Delta n_i\right) \mathbb{I}(n_{i-1}>0) \mathbb{I}(N+1 > n_i) \ .
\end{split}\eeq
Collecting all these definitions together, we can associate a Boltzmann weight $W(\boldsymbol{x}, \boldsymbol{n})$ 
with each possible alignment $(\boldsymbol{x}, \boldsymbol{n})$ of a given
sequence $\boldsymbol{A}$,
which takes into account all energetic contributions for feasible assignments only,
\beq\label{eq:Boltz-weight}
\begin{split}
W(\boldsymbol{x}, \boldsymbol{n}) & = \frac1Z e^{-\HH_{\rm DCA}(\boldsymbol{x}, \boldsymbol{n})} \chi_{\mathrm{in}}(x_1,n_1) \chi_{\mathrm{end}}(x_L,n_L)\\
& \prod_{i=2}^L\chi_{\mathrm{sr}}(x_{i-1},n_{i-1},x_{i},n_i)
 \prod_{i=1}^L\chi_{\mathrm{gap}}(x_i,n_i) \ ,
\end{split}\eeq
where $Z$ is the partition function.
Note that the ``hard constraints'' $\chi_{\rm{sr}}$, which can set the weight to zero, live only on the edges of the
linear chain $1,...,L$, while the interactions $J_{ij}$ are in principle fully connected. 

%

Finally, we can map the original minimization problem in
Eq.~\eqref{eq:Edef} as the statistical physics problem of finding the
best assignment of the variables $(\boldsymbol{x}^{*},
\boldsymbol{n}^{*})$ that maximizes the Boltzmann distribution:
\begin{equation}\label{eq:maxW}
  (\boldsymbol{x}^{*},\boldsymbol{n}^{*}) =
  \underset{(\boldsymbol{x},\boldsymbol{n})}{\mathrm{argmax}} \, W(\boldsymbol{x}, \boldsymbol{n}) \ .
\end{equation}
Alternatively, we could obtain an optimal alignment
$(\boldsymbol{x}^{*}, \boldsymbol{n}^{*})$ from an equilibrium
sampling of alignments with weight $W(\boldsymbol{x},
\boldsymbol{n})$.  Unfortunately, both sampling from
$W(\boldsymbol{x}, \boldsymbol{n})$ and identifying the constrained
optimal assignment are hard and intractable problems.  Note that the
space of possible assignments has dimension scaling as $(N+2)^L$,
which grows extremely quickly with $N$ and $L$. For comparison, the
DCA problem is defined in a space growing ``only'' as $q^L$.  However,
some approximations inspired by statistical physics can be exploited
for seeking an approximate solution.

\section{Advanced mean-field approximation}
\label{sec:advmf}

A straightforward approach to make this problem tractable is to use
message-passing approximations of the marginal probabilities
$P_{i}(x_{i}, n_{i})$ of Eq.~\eqref{eq:Boltz-weight}, such as Belief Propagation (BP), which are exact
for problems defined on graphs without loops. Note that BP is also
exact on linear chains, for which it coincides with the transfer
matrix method (or dynamic programming / forward-backward
algorithm \cite{durbin1998}). One can think to BP as 
treating exactly the linear chain $1,\cdots,L$, while the longer-range
interactions are approximated by message-passing.  In the case of
vanishing couplings $J_{ij}(A,B)\equiv0$ for all $i,j$ such that
$|i-j|>1$ and for all $A,B$, \ie in the case of a model with
nearest-neighbor interactions only, this formulation is exact; if all
couplings vanish, it is quite similar to a profile Hidden Markov
Model, which has the same penalties for opening and extending a
sequence of insertions.  However, our interactions are instead
typically very dense (all couplings are non-zero, hence the associated
graph is very loopy), but weak. We can
thus consider a further approximation of BP~\cite{decelle2011}, in which the linear chain
$1,\cdots,L$ is still treated exactly, while the contribution of more
distant sites is approximated via mean field (MF), in a way similar to
Thouless-Anderson-Palmer (TAP) equations~\cite{tap1977}, also known as Approximate
Message Passing (AMP) equations. We refer to this approach simply as
MF in the following.  In the rest of this section we derive the BP and
MF equations.

\subsection{Transfer matrix equations for the linear chain} 

Suppose first that only nearest-neighbor couplings $J_{i,i+1}(A,B)$
are non-zero. In this case, the problem is exactly 
solved by the transfer matrix method, which corresponds to a set of recursive equations for
the ``forward messages'' $F_i(x_i,n_i) = F_{i \to i+1}(x_i,n_i)$, \ie 
the probability distribution of site $i$ in absence of the link
$(i,i+1)$, 
and ``backward messages'' $B_i(x_i,n_i) = B_{i \to i-1}(x_i,n_i)$,
\ie the probability distribution of site $i$ in absence of the link
$(i-1,i)$.  

We give here the transfer matrix equations for the forward and backward messages. For compactness we define the single-site weight contribution
to Eq.~\eqref{eq:Boltz-weight}:
\begin{eqnarray}
\WW_{1}(x_{1},n_{1}) &=& \chi_{\rm{in}}(x_1,n_1) e^{ h_{1}(A_{x_{1}\cdot n_{1}})-(1-x_{1})\mu(n_{1})} \ , \nonumber \\
\WW_{i}(x_{i},n_{i}) &=&  e^{ h_{i}(A_{x_{i}\cdot n_{i}})-(1-x_{i})\mu(n_{i})} \ , \\
\WW_{L}(x_{L},n_{L}) &=& \chi_{\rm{end}}(x_{L},n_{L}) e^{ h_{L}(A_{x_{L}\cdot n_{L}})-(1-x_{L})\mu(n_{L})} \ , \nonumber
\end{eqnarray}
where the second line is for $i=2,\cdots,L-1$.
We then have for the forward messages:
\beq\label{eq:MF_F}
\begin{split}
&F_{1}(x_{1},n_{1})  =  \frac{1}{f_{1}}\WW_1(x_1,n_1) \ , \\ 
&F_{i}(x_{i},n_{i})  =  \frac{1}{f_{i}} \WW_{i}(x_{i},n_{i}) \FF_i(x_i,n_i) \ , \\
&\FF_i(x_i,n_i)  = \sum_{x_{i-1},n_{i-1}}F_{i-1}(x_{i-1},n_{i-1}) \times \\ 
&\times  e^{ J_{i-1,i}(A_{x_{i-1}\cdot n_{i-1}},A_{x_{i}\cdot n_{i}})} \chi_{\rm{sr}}(x_{i-1},n_{i-1},x_{i},n_i) \ , 
\end{split}\eeq
where $F_i$ is defined for $i=1,\cdots, L-1$ and
$\FF_i$ for $i=2,\cdots, L$, and the $f_i$ are normalization constants determined by the requirement that messages are normalized to one.
For the backward messages we have
\beq\label{eq:MF_B}
\begin{split}
&B_{L}(x_{L},n_{L})  =  \frac{1}{b_{L}} \WW_{L}(x_{L},n_{L})  \ , \\ 
&B_{i}(x_{i},n_{i})  = \frac{1}{b_{i}} \WW_{i}(x_{i},n_{i}) \BB_i(x_i,n_i) \ , \\
&\BB_i(x_i,n_i) = \sum_{x_{i+1},n_{i+1}}B_{i+1}(x_{i+1},n_{i+1}) \times \\ 
& \times e^{ J_{i,i+1}(A_{x_{i}\cdot n_{i}},A_{x_{i+1}\cdot n_{i+1}})} \chi_{\rm{sr}}(x_{i},n_{i},x_{i+1},n_{i+1}) \ ,
\end{split}\eeq
where $B_i$ is defined for $i=L,L-1,\cdots,2$ and $\BB_i$ for $i=L-1, \cdots, 1$, and $b_i$ are normalization constants.
Finally, the marginal probabilities are given by
\begin{eqnarray}
P_{1}(x_{1},n_{1}) & = & \frac{1}{z_{1}} \WW_1(x_1,n_1) \BB_1(x_1,n_1)  \ , \nonumber \\
P_{i}(x_{i},n_{i}) & = & \frac{1}{z_{i}} \WW_i(x_i,n_i) \FF_i(x_i,n_i)\BB_i(x_i,n_i)  \ , \\
P_L(x_L,n_L) &=& \frac{1}{z_{L}} \WW_L(x_L,n_L) \FF_L(x_L,n_L) \ . \nonumber
\end{eqnarray}
These equations are summarized in compact form in Table~\ref{tab:1}.
They can be easily implemented on a computer and solved in a time scaling as $L \, N$.

\begin{table}[t]
  \begin{tabular}{c|l|c}
    $i=1$ & $i=2,\cdots,L-1$ & $i=L$ \\
  \hline
  $P_{1}  = \frac{1}{z_{1}} \WW_{1}  {\cal B}_1$  
  & $P_{i}  = \frac{1}{z_{i}} \WW_{i} {\cal F}_i {\cal B}_i$ 
  &$P_L = \frac{1}{z_L} \WW_L {\cal F}_L$ 
  \\
  \hline
  $F_1  = \frac{1}{f_{1}}\WW_{1}$  
  & $F_i  = \frac{1}{f_{i}} \WW_i{\cal F}_i$  
  & $\text{---}$ \\
  \hline
  $\text{---} $
  & $B_i  = \frac{1}{b_{i}} \WW_i {\cal B}_i$ 
  &$B_L = \frac{1}{b_{L}} \WW_L$ 
\end{tabular}
\caption{Schematic summary of the transfer matrix and mean field equations, which are complemented by the recurrence equations
for $\FF_i$ and $\BB_i$ given in Eqs.~\eqref{eq:MF_F} and \eqref{eq:MF_B}.
For mean field one should replace
$\WW_i \to \CC_i$.}
\label{tab:1}
\end{table}

\subsection{Long range interactions} 
\label{seq:eqs}

We now discuss the inclusion of long-range interactions in the transfer matrix scheme.
In order to treat correctly the long-range interaction in BP, it is important
to note that the same ``light-cone'' condition expressed by the
constraint $\chi_{\rm{sr}}$ in Eq.~\eqref{eq:cond} 
holds between any pair $(i,j)$. However, this condition would be
violated by the messages of BP due to their approximate character on loopy graphs. In
order to enforce it, we can introduce a new constraint
\begin{eqnarray}
\label{Boltz}
\chi_{\rm{lr}}&&\left(x_{i},n_{i},x_{j},n_{j}\right)  = \\  
&& \mathbb{I}\left[i>j+1\right]\left\{ \delta_{x_{i},0}\mathbb{I}\left[n_{i}\geq n_{j}\right]  + \delta_{x_{i},1}\mathbb{I}\left[n_{i}>n_{j}\right]\right\} \nonumber   \\
+&&\mathbb{I}\left[i<j-1\right]\left\{\delta_{x_{j},0}\mathbb{I}\left[n_{i}\leq n_{j}\right]+\delta_{x_{j},1}\mathbb{I}\left[n_{i}<n_{j}\right]\right\} \ . \nonumber
\end{eqnarray}
Because this constraint is redundant with respect to Eq.~\eqref{eq:cond}, it can be added without changing the weight:
\begin{equation}
W(\boldsymbol{x},\boldsymbol{n}) = W(\boldsymbol{x},\boldsymbol{n}) \times \prod_{i<j} \chi_{\rm{lr}}\left(x_{i},n_{i},x_{j},n_{j}\right) \ .
\end{equation}
However, adding this constraint ensures that the proper ordering of the pointers is preserved under the BP approximation.

The BP equations can be written straightforwardly (see SI) and provide
an approximation to the marginal probabilities $P_i$. 
A naive implementation requires a time scaling
as $L^3 N^2$, which can be easily reduced to $L^2 N^2$. {\tiny {\tiny }}In the SI we discuss a
simplification of the BP equations, under the assumption that pairs of
sites with $|i-j|>1$ can be treated in a mean field~\cite{decelle2011}.  We find that the
resulting mean field equations are identical to the transfer matrix
ones, with the only replacement of the local weight $\WW_i \to \CC_i$,
with
\begin{widetext}
\beq
{\cal C}_{i}(x_{i},n_{i})  =  \WW_i(x_i,n_i) \, \exp\left\{\sum_{j\notin\{i,i\pm1\}}\sum_{x_{j},n_{j}}\chi_{\rm{lr}}(x_{i},n_i,x_j,n_{j})J_{i,j}(A_{x_{i}\cdot n_{i}},A_{x_{j}\cdot n_{j}})P_{j}(x_{j},n_{j}) \right\}  \ .
\eeq
\end{widetext}
As a result, the mean field equations have the same complexity as the
transfer matrix equations $(L N)$ with an additional factor $L N$
needed to compute each $\CC_i$, resulting in an overall complexity
$L^2 N^2$, the same scaling of the BP update. However, while the memory consumption needed by BP to store all the incoming messages at each node scales as $2 \cdot L^2 \cdot N$, MF has the advantage of working directly on the $ 2 \cdot L \cdot N $ approximated marginal probabilities.

\subsection{Assignment}
\label{sec:assignment}

After solution of the MF equations, from the marginal probabilities
$\{P_{1}(x_1,n_1), ...,P_{L}(x_L,n_L)\}$ we have to find the most
probable assignment $(\boldsymbol{x}^{*}, \boldsymbol{n}^{*})$, as
defined in Eq.~\eqref{eq:maxW}.  The simplest way to do so is to
assign to each position $i$ the most probable state according to its
marginal, \ie
\begin{equation}
\left(x^{*}_{i},n^{*}_{i}\right) = \underset{x_{i},n_{i}}{\text{argmax}}\,P_{i}\left(x_{i}, n_{i}\right) \ ,
\end{equation}
which is possible whenever the obtained assignment satisfies all the
hard constraints.  However, in some cases, the set of locally optimal
positions do not satisfy the short-range constraints due to the
approximate nature of the MF solution.  We then perform a
\textit{maximization} step, in which we select the position $i^{*}$
and the local assignment $(x^{*}_{i^{*}},n^{*}_{i^{*}})$ having the
largest probability among all the marginals, \ie
\beq
(i^*,x^{*}_{i^{*}}, n^{*}_{i^{*}}) = \underset{i,x_i,n_i}{\text{argmax}} \{P_{1}(x_1,n_1),...,P_{L}(x_L,n_L)\} \ .
\eeq
We then set the state of site $i^{*}$ in
$(x^{*}_{i^{*}},n^{*}_{i^{*}})$, and we proceed with a
\textit{filtering} step, in which we set to zero the marginal
probabilities of the incompatible states of the first nearest
neighbors of $i^*$.  In practice, we multiply the marginals by the
short-range constraints computed at $(x^{*}_{i^{*}}, n^{*}_{i^{*}})$,
\ie we consider the new marginals on sites $i^*\pm 1$:
\begin{equation}
  \begin{split} &P_{i^{*}-1} (x_{i^{*}-1}, n_{i^{*}-1})
  \chi_{\rm{sr}}(x_{i^{*}-1},n_{i^{*}-1},n^{*}_{i^{*}},x^{*}_{i^{*}})
  \ , \\ &P_{i^{*}+1} (x_{i^{*}+1}, n_{i^{*}+1})
  \chi_{\rm{sr}}(x^{*}_{i^{*}},n^{*}_{i^{*}},x_{i^{*}+1},n_{i^{*}+1})
  \ .
  \end{split}
\end{equation}
We can now repeat the \textit{maximization} step in order to find the
state $(x^{*},n^{*})$ that maximizes the joint set of probabilities
for the positions adjacent to the already aligned part of the
sequence (in this case $i^{*}-1$ and $i^{*}+1$, because we only fixed
$i^{*}$),
\begin{eqnarray}
(x^{*},n^{*}) = \underset{x,n}{\text{argmax}} \{P_{i^{*}-1} (x,n),P_{i^{*}+1} (x,n)\} \ ,
\end{eqnarray}
and we fix this state, in the alignment, in the right position (either $i^{*}+1$ or $i^{*}-1$). Suppose for simplicity that we have just specified the state in position $i^{*}+1$; we now filter the probability of $i^{*}+2$ and repeat the choice for the next $(x^{*},n^{*})$ considering the set of (modified) marginals $\{P_{i^{*}-1}(x,n), P_{i^{*}+2}(x,n)\}$. 
The procedure is repeated until all the $L$ positions are determined. 

Note that this scheme is somehow greedy, because the assignments are
decided step by step, and are constrained by the choices made in the
previous positions. Still, the assignment is guided by the marginal
probabilities obtained from considering the global energy function and
all the hard constraints. Moreover, this assignment procedure is as
fast as the ``max-marginals" scheme because it does not require to
re-run the update of the equations in Sec.~\ref{sec:advmf}, and thanks
to the step-by-step filtering of the marginals it ensures an outcome
that is always compatible with the constraints.

\subsection{Discussion}

In this section we presented a set of approximate equations and an
assignment procedure that, together, allow us to solve the alignment
problem in polynomial time. In both BP and MF, the equations can be solved at
``temperature'' equal to one, corresponding to a Boltzmann equilibrium
sampling from the weight in Eq.~\eqref{eq:Boltz-weight}, or at zero
temperature, corresponding to finding (approximately) the most likely
assignment in Eq.~\eqref{eq:Boltz-weight} (the full set of equations for MF
at zero temperature is reported in the SI). Of course, any
intermediate temperature could also be considered, but we do not
explore other values of temperature in this work.

In all cases,
one can compute the free energy associated with the BP or MF solution, which gives a ``score'' measuring the quality of the alignment. This score could be used, in long
sequences with multiple hits, to decide a ``best hit''. The expression
of the free energy is given in the SI. 

The MF equations are derived from the BP equations by assuming that all couplings with $|i-j|>1$ are weak enough to be treated in mean field. But we know that in (good) 
protein models, stronger couplings correspond to physical contacts in the three-dimensional structure, while a background of weaker couplings describe other correlations 
or even just noise. It could be interesting, therefore, to use a mixed BP/MF method, in which weaker couplings with $||J_{ij}||<K$ are treated in mean field, while 
stronger couplings with $||J_{ij}||\geq K$ are treated with BP, for a given threshold $K$. 
The case $K=0$ corresponds to pure BP, while the case $K\to\io$ corresponds to pure MF. One could check 
whether an optimal value of $K$ exists, but we leave this for future work.
We show below the results obtained using MF, which seems preferable to BP in the cases we analyzed. 
The MF equations converge faster than BP to their fixed point and, additionally, BP is more affected by the loopy character of the graph,
and often converges to local minima of the energy landscape. We leave
a more systematic comparison of the BP and MF schemes 
to future developments.

\section{Learning the model}
\label{sec:learning}

We now discuss the learning of the model parameters, namely the
couplings and fields of the DCA Hamiltonian, and the hyper-parameters
$\boldsymbol{\mu}$, $\boldsymbol{\lambda}$.

\subsection{Potts model}
\label{sec:models}

Our alignment method is able to cope with different cost functions,
because the implementation of the update equations described in
Sec.~\ref{seq:eqs} is as general as possible. Introducing a 5-state
alphabet, the method is also able to treat RNA alignments.

In this work, we tested several types of Maximum Entropy models for
DCA, which differ in the choice of fitted observables.  The usual
Potts model, in which all first and second moments of the seed MSA are
fitted, is labeled as \textit{potts}, while we also consider a
``pseudo" Hidden Markov Model (\textit{phmm}), with Hamiltonian
\begin{equation}
  \begin{split}
    &\mathcal{H}_{phmm}(\boldsymbol{S} | \boldsymbol{J}, \boldsymbol{h}) =\\
    & =- \sum_{i,i+1}J_{i,i+1} \delta_{S_i,-} \delta_{S_{i+1},-} - \sum_{i}h_{i}(S_{i}) \ .
  \end{split}
\end{equation}
The {\it phmm} can be thought of as a profile model playing the role
of the emission probabilities of a Hidden Markov Model (HMM), plus a
pairwise interactions $J_{i,i+1}$ between neighboring gaps. This
interaction is related, in our mapping to a HMM, to the probability of
switching, between two consecutive positions, from a ``gap'' to a
``match'' state and vice-versa.  We also considered other variations of
the Potts model, such as a model in which we do not fit the second
moment statistics of non-neighboring gaps (\ie long-range gap-gap
couplings are set to zero). The motivation behind this choice is that,
if DCA couplings are interpreted as predictors for the (conserved)
three dimensional structure, gapped states do not carry any
information about co-evolution of far-away positions.  However, we
found that these other variations do not bring additional insight with
respect to the \textit{potts} and \textit{phmm} models, so we restrict
here the presentation to these two choices.

All these models are learned on a seed alignment using a standard
Boltzmann machine DCA learning algorithm~\cite{figliuzzi2018pairwise}.
We used a constant learning rate of
$5\cdot10^{-2}$ for most protein families, and $10^{-2}$ for all RNA
families and for the longest protein families we used.  Because the
seed often contains very few sequences, we need to introduce a small
pseudo-count of $10^{-5}$ to take into account non-observed empirical
second moments. The Boltzmann machine performs a Monte Carlo sampling
of the model using $1000$ independent chains and sampling $50$ points
for each chain (in total the statistics is thus computed using
$5\cdot10^{4}$ samples). Equilibration and auto-correlation tests are
performed to increase or decrease, if needed, the equilibration or the
sampling time of the Monte Carlo. 

It is important to keep in mind that the models inferred from the
seeds are ``non-generative" because of the reduced number of sequences
(samples generated from these models, due to a strong over-fitting,
are extremely close to seed sequences), but nonetheless they are
accurate enough to be used as proper cost functions for our alignment
tool. We also mention that models inferred from Pseudo-Likelihood
maximization \cite{aurell2013,ekeberg2014}, which are also known to be
non-generative \cite{figliuzzi2018pairwise}, can be equivalently used
for the alignment method described in this work.

\subsection{Insertion penalties} \label{sec:MLinspen}

\begin{figure*}
	\includegraphics[width=1.0\textwidth]{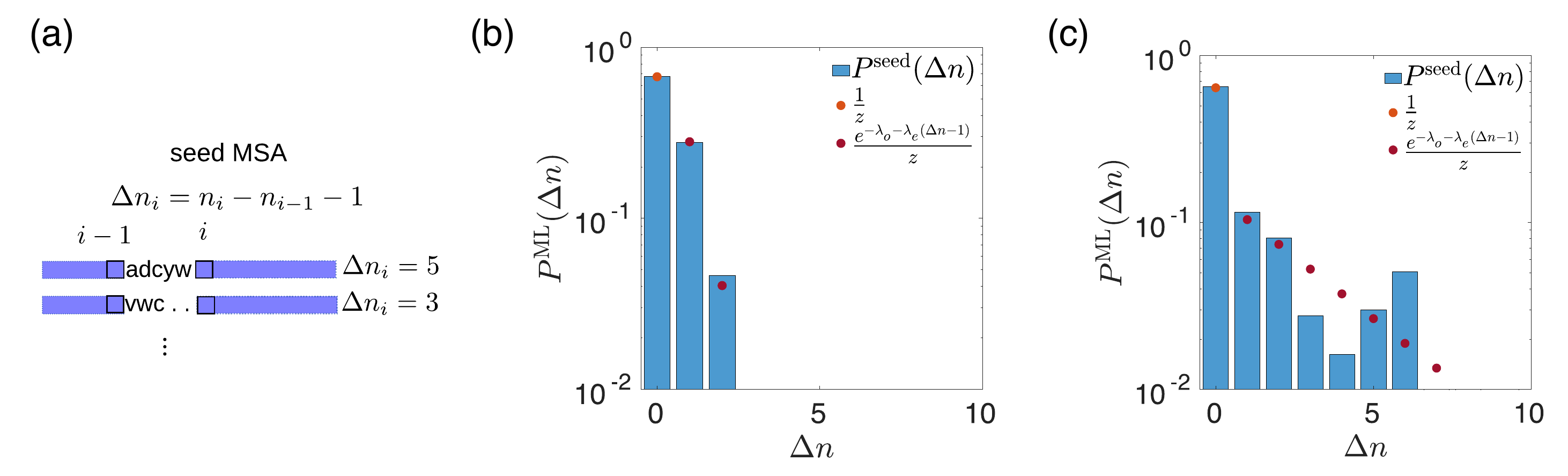}
	\caption{\textbf{Inference of Insertion penalties}. (a)
          Schematic representation of the $\Delta n$ variables: the
          number of insertions between two consecutive positions in
          the alignment can be computed from the pointer variables
          $\boldsymbol{n}$. (b),(c) Examples of fitting of the
          empirical probability of $\D n$ using our maximum likelihood
          approach for (b) position 25 and (c) position 92 for
          RF00162. In (c) the data distribution does not show an
          exponential profile but our approximation fits well the
          empirical probability for most of the observed $\Delta
          n$.  \label{fig:lambda}}
\end{figure*}

We determine the parameters of the affine insertion penalties using
the statistics of the insertions of the seed alignment.  Recall that
the number of insertions between positions $i$ and $i-1$ is $\Delta
n=n_{i}-n_{i-1}-1$, as illustrated in Fig.~\ref{fig:lambda}.
Motivated by the empirical statistics of insertions in true seeds, we
model the probability of $\Delta n$ as
\begin{eqnarray}
P_{i}\left(\Delta n\right)	=	\begin{cases}
	\frac{1}{z} \ ,& \Delta n=0 \ ,\\
	\frac{e^{-\lambda^{i}_{o}-\lambda^{i}_{e}\left(\Delta n-1\right)}}{z} \ , & \Delta n>0 \ ,
\end{cases}
\end{eqnarray}
where 
\beq\label{eq:z} 
z	 =	 1 + \sum_{\Delta n>0}e^{-\lambda^i_{o}-\lambda^i_{e}\left(\Delta n-1\right)} 
 =	1 + \frac{e^{-\lambda^i_{o}}}{1-e^{-\lambda^i_{e}}} 
 \eeq
is the normalization constant and $\lambda^{i}_{o}$, $\lambda^{i}_{e}$
are the costs associated with the opening and the extension of an
insertion as in the score function defined in Eq.~\eqref{eq:HHins}.
Because the learning of the parameters is done independently for each
position $i$, for the sake of simplicity we will drop the index $i$ in
the following.

We determine the values of $\lambda_{o}$, $\lambda_{e}$ by maximizing
the likelihood $\mathcal{L}(\left\{ \Delta n\right\}
_{a=1}^{M}\mid\lambda_{o},\lambda_{e})$ of the data, \ie the $M$
sequences of the seed, given the parameters, and adding
L2-regularization terms in order to avoid infinite or undetermined
parameters.  Imposing the zero-gradient condition on the likelihood
leads to a closed set of non-linear equations for the maximum
likelihood estimators, given in the SI.  These equations can be
solved, for example, by a gradient ascent scheme in which we iteratively
update
\beq \lambda^{t+1} \leftarrow
\lambda^{t}+\eta\frac{\partial\mathcal{L}_{t}}{\partial\lambda} \ ,
\eeq
for both $\lambda_{o}$ and $\lambda_{e}$, until numerical convergence
(more precisely, when the absolute value of the gradient is less then
$10^{-4}$). The learning rate is $\eta=10^{-3}$.  Note that the
empirical distribution can differ from that of our model: for
instance, we often encounter positions where either no insertion is
present within the seed or the distribution of the positive $\Delta n$
is not exponential. In the first case, our maximum likelihood approach
cannot be applied as it is: in order to apply it we pretend that the
probability of observing at least one insertion is equal to a small
parameter $\epsilon$ so that we can slightly modify
$P^{\rm{seed}}(\Delta n = 0) = 1 - \epsilon$. In our work this
parameter has been set to $\epsilon = 10^{-3}$. In the second case we
notice that the distribution given by the fit is anyway a nice
approximation of the true one. Some examples are reported in
Fig.~\ref{fig:lambda}.

\subsection{Gap penalties}
\label{sec:supervgaps}

The gap state is treated in DCA models as an additional amino acid
but, by construction of the MSA, it is actually an \textit{ad hoc}
symbol used to fill the vacant positions between well-aligned
amino acids that are close in the full-length sequence
$\boldsymbol{A}$ and should be more distant in the aligned sequence
$\boldsymbol{S}$. Thus the proper number of gaps for each candidate
alignment is often sequence dependent and not family dependent: the
one-point and two-points statistics of gaps computed from the seed may
not be representative of the full alignment statistics. Yet, the
couplings and the fields of the DCA models learned from the seed tend
to place gaps in the positions mostly occupied by gaps in the
seed. This may lead to some bias depending on the seed construction:
we notice that if we create seeds using randomly chosen subsets of
Pfam \cite{el2019pfam} alignments produced by HMMer
\cite{eddy2011accelerated}, our alignment method, DCAlign, is likely to 
produce very gapped sequences. In these cases gaps appear very often,
more often than any other amino acid, indicating that our cost function
encourages the presence of gaps.  Real seeds are instead manually
curated and therefore they generally contain few gaps. Even though
the Potts models learned from this kind of seeds are less
biased, we anyway need to check whether the issue exists and, if
needed, treat it. To do so, our idea is to introduce additional
penalties to gap states, $\mu_{\rm ext}$ and $\mu_{\rm{int}}$, as we
discussed in Sec.~\ref{sec:penalties} in the definition of the cost
function.  Notice that the distinction between ``internal" and
``external" gap penalties allows us to differentiate between gaps that
are artificially introduced (as in the case of the internal ones) and
gaps that reflect the presence of well aligned but shorter domains
or fragments, of effective length $L_{\rm{frag}} < L$. In this last
case some ``external" gaps are needed to fill the $L-L_{\rm{frag}}$
positions at the beginning or at the end of the aligned sequence.

Contrarily to the insertions penalties, the gap penalties cannot be
directly learned from the seed alignment via an unsupervised training
(as their statistics is already included in the Potts model to begin
with), but they can be learned in a supervised way.  A straightforward
procedure consists in re-aligning the seed sequences using the
insertions penalties and the DCA models (\textit{potts} or
\textit{phmm}) described in Sec.~\ref{sec:models} for several values
of $\mu_{\rm ext}$ and $\mu_{\rm{int}}$. The best values of the gap
parameters are those that minimize the average Hamming distance
between the re-aligned seed and the original seed sequences.  We
performed this supervised learning by setting the values of $\mu_{\rm
  ext} \in [0.00,\,4.00]$ and $\mu_{\rm{int}} \in [0.00,\,
  4.00]$. These intervals have been chosen after several tests in a
larger range of variability, also including negative values (that
favor gaps) and very large values compared to the typical parameters
of the Potts models.  We observed that: (i)~favoring gaps is always
counterproductive, and (ii)~there exists a threshold, usually around 4,
beyond which no gap is allowed in the sequence, which is also
counterproductive. For these reasons, and because of the high
computational effort required to re-align the sequences several times,
we decided to use the interval $\left[0.00, 4.00\right]$ with
sensitivity 0.50 leading to 81 re-alignments of the seed sequences.

This method works for seeds that contain a large number of sequences
(typically $M>10^{3}$) but it fails completely when dealing with
``small" seeds. In this case, whatever the value of ($ \mu_{\rm
  ext},\mu_{\rm{int}}$), the re-aligned sequences are always identical
to the original ones, resulting in an average Hamming distance equal
to zero. Indeed, the energy landscape of models learned from few
sequences is populated by very isolated and deep local minima centered
in the seed sequences. When the algorithm tries to re-align an element
of the training set, it is able to perfectly minimize the local energy
and re-align the sequence with no error whatever the additional gap
penalty.  For short seeds, instead of re-aligning the seed, we thus
extract $1500$ sequences from the full set of unaligned sequences (a
\textit{validation} set), which we align by varying the gap penalties,
always in the range $[0.00,\,4.00]$. We call
$\HH^{\rm{0}}_{\rm{seed}}$ the DCA Hamiltonian inferred on the seed,
and we infer new Hamiltonians $\HH^{\rm{0}}_{type,\mu_{\rm
    ext},\mu_{\rm{int}}}$ (with ${type \in \{phmm, potts,\}}$) on all
the multiple sequence alignments of the validation set.  We then
choose the best parameters $\mu_{\rm ext}$ and $\mu_{\rm int}$ as
those that minimize the symmetric Kullback-Leibler distance between
$\HH^{0}_{seed}$ and $\HH^{\rm{0}}_{type,\mu_{\rm
    ext},\mu_{\rm{int}}}$ (the precise definition is given in
section~\ref{sec:exps}). In other words, we select the best gap
penalties as those that produce a validation MSA as statistically
close as possible to the seed alignment.

We underline that the values of the penalty parameters also depend on
the choice of the gauge for the DCA parameters: in fact, the advanced
mean-field equations are not gauge invariant and depending on the
choice of the gauge we can have different (optimal) values for the gap
penalties. All results shown in this work use the zero-sum gauge for
the DCA parameters.

\section{Computational setup}
\label{sec:exps}

\subsection{Pipeline} 
\label{sec:pipeline}

\begin{figure*}[t]
	\centering
	\includegraphics[width=1\textwidth]{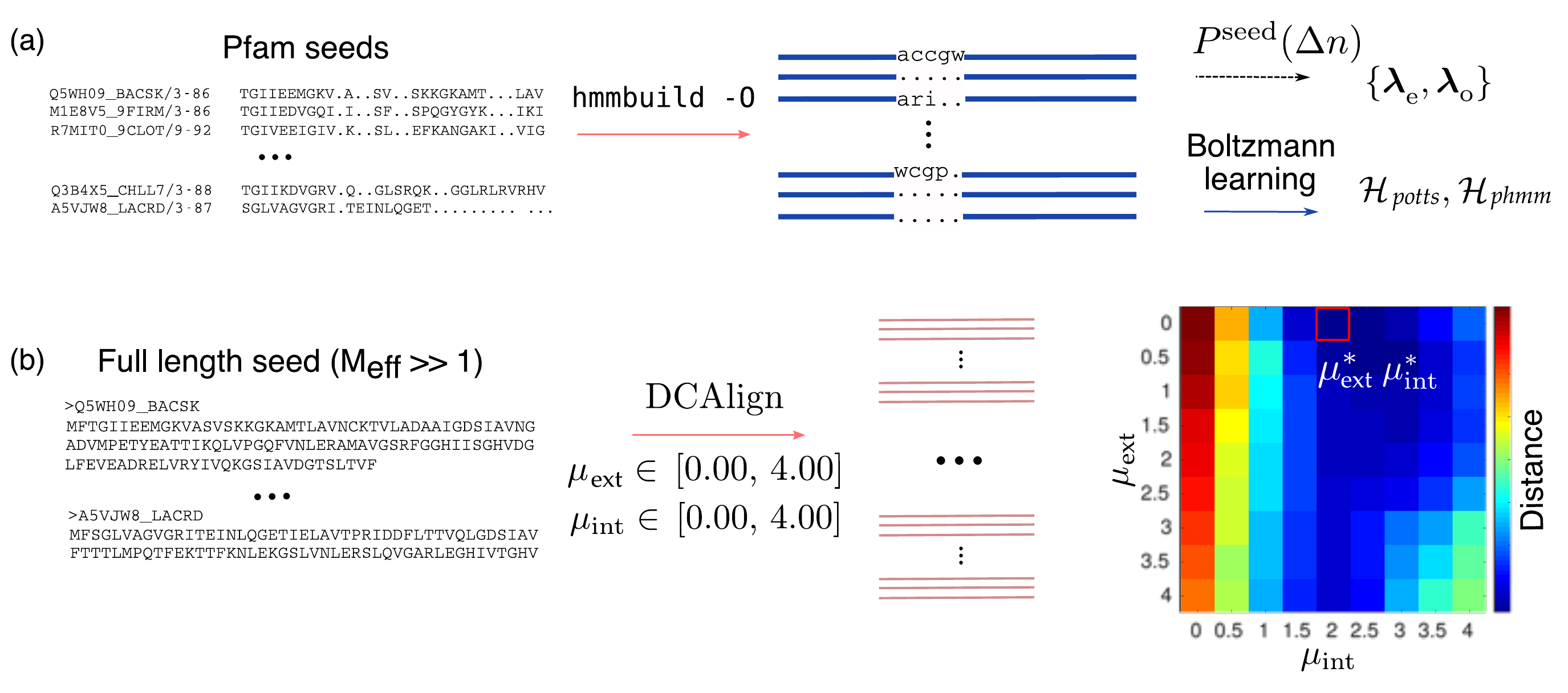}
	\caption{\textbf{Scheme of the training process for DCAlign.}
          In (a) we show the first step of the learning. 
          We build the aligned seed of a Pfam family using
          $\texttt{hmmbuild -O}$ to detect the matched amino-acids
          (blue line) and the insertions (shown in lower case letters
          and ``."). From these data we learn the DCA Hamiltonian and
          the affine insertion penalties. In (b) we pictorially
          describe the learning of the gap penalties. Here we take
          into account the seed itself (not only the aligned part but
          the entire sequences) and we try to align it using the
          parameters inferred in step (a) using all the 81 possible
          combinations of $\mu_{\rm ext}$ and $\mu_{\rm{int}}$, each
          spanning the interval $\left[0.00, 4.00\right]$. We compare
          each candidate alignment to the seed alignment, directly
          using the Hamming distance. The best set of parameters is
          that minimizing this metric. The plot in (b) shows the
          (average) Hamming distance of the true and re-aligned seed
          for the PF00677 family.\label{fig:train}}
\end{figure*}

The computational setting we propose here is the same adopted by
state-of-the-art alignment softwares such as HMMer \cite{hmmerweb},
and Infernal \cite{nawrocki2013infernal}. From
the seed alignment we learn all the parameters that characterize our
score function and we apply our alignment tool to all the unaligned
sequences that contain a domain compatible with the chosen family.
More precisely, once a seed is selected (we used the
  \texttt{hmmbuild -O} function of the package HMMer to obtain the
  aligned seed), we learn the model, the insertions parameters and
the gap penalties as described in Sec.~\ref{sec:learning}.  A scheme
of our training method is shown in Fig.~\ref{fig:train}.

After training, our cost function is fully determined. Unaligned
sequences are then taken from the full length sequences of the
corresponding family in the Pfam database \cite{el2019pfam} (we do not
face here the 
problem of detecting homologous sequences). Note that, like HMMer, we
also include the seed sequences in the sequences to be aligned, in
order to obtain a more homogeneous MSA and test the quality of the
re-alignment of the seed.  We do not consider the entire sequences,
whose length $N$ is often much larger than $L$, but a ``neighborhood"
of the hit selected by HMMer. In practice we add 20 amino-acids at the
beginning and at the end of the hit resulting in a final length $N =
20 + L + 20$. We performed the same experiment using $N = 50
  + L + 50$ for PF00684 and the resulting sequences were identical to
  those obtained from a shorter hit. The method seems to be stable for
  reasonable values of $N$, \ie $N \sim L$.  For RNA sequences, this
pre-processing is not needed, because the full length sequences
downloaded from Rfam already have a reasonable length. For most
families, we have aligned all the full length sequences (the size of
the test sets is specified in Table \ref{tab:info_fam}) and only in
few cases, for particularly large families, we uniformly pick at
random $N_{\rm{seq}} = 10^4$ sequences to align.

We then apply DCAlign using the approximations we discussed in
Sec.~\ref{sec:advmf}, namely the finite temperature and
zero-temperature MF method, to each candidate sequence and we add to
our MSA the aligned sub-sequence that has the
lower energy (insertion and gap penalties excluded).  Whenever our
algorithms do not converge to an assignment of the variables that
satisfies all the hard constraints, we apply the ``nucleation" procedure
explained in Sec.~\ref{sec:assignment} that, by means of the
approximated marginal probabilities, gives rise to feasible
alignments.

\subsection{Observables}
\label{sec:observables}

To assess the quality of the MSAs generated by DCAlign (that differ
in the score function being used to align) and to compare them to the
state-of-the-art alignments provided by HMMer (or Infernal), we
consider the following observables.

\begin{itemize}

\item{\textit{Sequence-based metrics.}} When comparing two candidate MSAs of the same set of sequences (a ``reference'' and a ``target''), it is possible to compute several sequence-wise measures such as the following metrics (normalized by $L$, the length of the sequences):
	\begin{itemize}
	\item the Hamming distance between the two alignments of the
          same sequence in the reference and target MSAs;
	\item ${\rm Gap_{+}}$: the number of match states in the aligned sequence of the reference MSA that have been replaced by a gap in the target MSA;
	\item ${\rm Gap_{-}}$: the number of gaps in the aligned sequence of the reference MSA that have been replaced by match states in the target MSA;
	\item ${\rm Mismatch}$: the number of amino-acid mismatches,
          that is the number of times we have match states in both
          sequences, reference and target, but to different
          amino acids (positions) in the full sequence
          $\boldsymbol{A}$.
        \end{itemize}

\item{\textit{Proximity measure.}} Consider two different MSAs of the same protein or RNA family.
We can compute, for each candidate sequence $S_{i}^{1}$ of the first set, the Hamming distance $d_{H}$ with respect to all the sequences of the second set. 
We then collect the minimum attained value, \ie
\begin{equation}
	\hat d_i = \min_{j} d_{H}(S^{1}_{i}, S^{2}_{j}) \ ,
	\label{eq:proximity}
\end{equation}
which gives the distance to the closest sequence in the other MSA.
The distribution of $\hat d_i$, or some statistical quantity computed
from them (such as the average or the median value) provides a good
measure of ``proximity" between the two sets.  We will show below a
few examples using the full alignment of a protein
family as a first set, and the seed sequences as the second one.

\begin{figure*}[t]
	\centering
	\includegraphics[width=1.0\textwidth]{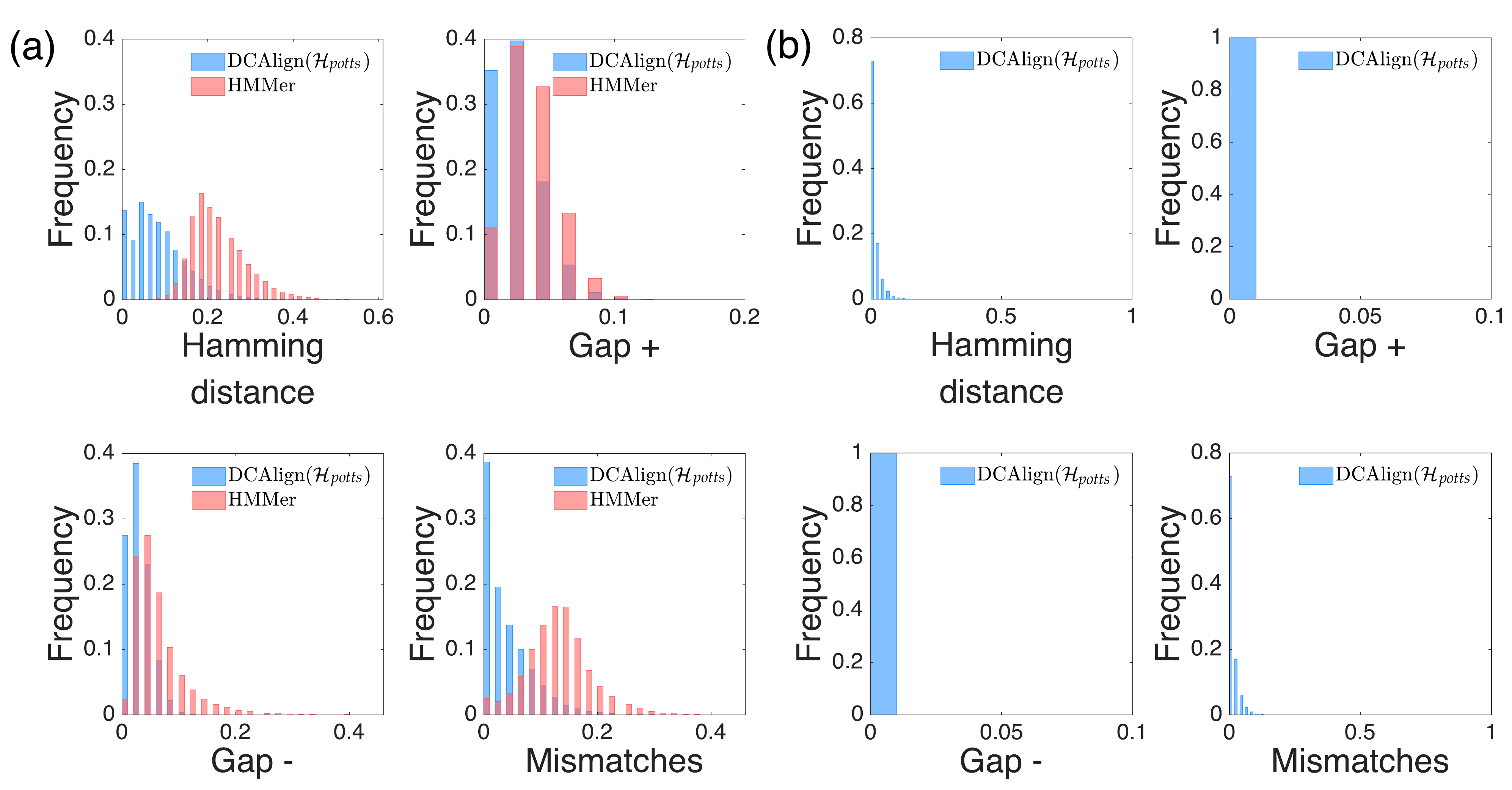}
	\caption{\textbf{Comparison between DCAlign and HMMer for
            synthetic data.} Panels (a) and (b) show the histograms
          of the normalized metrics (Hamming distance, Gap$_+$,
          Gap$_-$ and Mismatches), respectively, in the case of
          conserved sited only, and of correlated pairwise columns
          only. Here, the reference is the ground truth and the target
          is the alignment of DCAlign (blue) or HMMer (red).  In (b)
          HMMer results are not shown because $\texttt{hmmsearch}$
          does not find any relevant hit. \label{fig:metrics_synth}}
\end{figure*}

\item{\textit{Symmetric Kullback-Leibler distance}}. Another convenient
  global measure is the symmetric Kullback-Leibler distance between a
  Boltzmann equilibrium model learned from the seed alignment (the
  ``seed'' model) and another model learned from a candidate MSA (the
  ``test'' model). In general the symmetric Kullback-Leibler
  divergence is a measure of ``distance" between two probability
  distributions and it is defined, for arbitrary densities
  $\mathcal{P}_{1}(\boldsymbol{x})$ and
  $\mathcal{P}_{2}(\boldsymbol{x})$ of the variables $\boldsymbol{x}$,
  as
\begin{equation}
D_{{\rm KL}}^{{\rm sym}}\left(\mathcal{P}_{1},\mathcal{P}_{2}\right)=D_{{\rm KL}}^{{\rm }}\left(\mathcal{P}_{1}||\mathcal{P}_{2}\right)+D_{{\rm KL}}^{{\rm }}\left(\mathcal{P}_{2}||\mathcal{P}_{1}\right) \ ,
\end{equation}
where $D_{\rm{KL}}$ is the Kullback-Leibler divergence
\begin{equation}
D_{\rm{KL}} \left(\mathcal{P}_{1}||\mathcal{P}_{2}\right) = \sum_{x} \mathcal{P}_{1}(\boldsymbol{x}) \log \frac{\mathcal{P}_{1}(\boldsymbol{x})}{\mathcal{P}_{2}(\boldsymbol{x})} \ .
\end{equation} 
In our context, the symmetric KL distance can be efficiently computed through averages of energy differences as
\begin{equation}\begin{split}
&D_{{\rm KL}}^{{\rm sym}}\left(\mathcal{P}_{{\rm seed}},\mathcal{P}_{{\rm test}}\right) = \\
&=D_{{\rm KL}}^{{\rm }}\left(\mathcal{P}_{{\rm test}}||\mathcal{P}_{{\rm seed}}\right)+D_{{\rm KL}}^{{\rm }}\left(\mathcal{P}_{{\rm seed}}||\mathcal{P}_{{\rm test}}\right) \\ &=
\left\langle \mathcal{H}_{{\rm seed}}-\mathcal{H}_{{\rm test}}\right\rangle _{\mathcal{P}_{{\rm test}}}+\left\langle \mathcal{H}_{{\rm test}}-\mathcal{H}_{{\rm seed}}\right\rangle _{\mathcal{P}_{{\rm seed}}} \ ,
\end{split}\end{equation}
where $\mathcal{P}_{\left(.\right)}$ is the Boltzmann distribution
associated with the energy $\mathcal{H}_{\left(.\right)}$, and the
brackets $\left\langle...\right\rangle_{\mathcal P}$ denote the
expectations with respect to $\mathcal P$, which can be easily
estimated using a Monte Carlo sampling (in contrast to the normal
$D_{\rm KL}$, which depends on the intractable normalization constants,
\ie the partition functions,
of the densities $\mathcal{P}_{\left(.\right)}$). We run the
comparison using two different models for $\HH$, that is a Potts model
and a profile model.

\item{\textit{Contact map}}. The couplings of DCA models can be used
  to detect the presence of physical interactions between pairs of
  sites, which are distant in the one-dimensional chain but in close
  contact in the three-dimensional structure. A good score
  that indicates a direct contact is the (average-product corrected)
  Frobenius norm of the coupling matrices, defined as
\begin{eqnarray}
\mathcal{F}^{APC}_{i,j} = \mathcal{F}_{i,j} - \frac{\sum_{m}\mathcal{F}_{i,m} \sum_{n}\mathcal{F}_{n,j}}{\sum_{m,n}\mathcal{F}_{m,n}} \ ,
\end{eqnarray} 
where 
\begin{eqnarray}
\mathcal{F}_{i,j} & = & \sqrt{\sum_{A\neq'-',B\neq'-'}J_{i,j}(A,B)^{2}}  \ ,
\end{eqnarray} 
and the couplings are in the zero-sum gauge.  We thus compare
predicted contact maps obtained using the parameters learned from our
alignments and those obtained by HMMer. For this purpose, we use the
PlmDCA method to learn the couplings from the alignments, because it
is faster than the Boltzmann machine and it is known to be reliable
for contact prediction \cite{aurell2013}. The ground-truths denoting
the physical interactions in each protein are obtained running the
{\it Pfam interactions} package \cite{pfam_inter}. Two sites are said
to be in contact if the minimum atomic distance among all the atoms
and among all the available protein structures is less than 8
$\mathring{A}$.
\end{itemize}


\section{Test on synthetic data}

Here we describe two experiments on synthetic data, constructed to
compare the performances of DCAlign to state-of-the-art methods in
extreme settings: one dataset presents conserved but not coevolving
sites (\ie strong variations in amino-acid frequency on each site but
no correlation between distinct sites), while the other presents not
conserved but coevolving sites (\ie uniform frequency $1/q$ of
amino-acids on each site, but strong correlations between distinct
sites).

\begin{figure}[t]
  \includegraphics[width=.95\columnwidth]{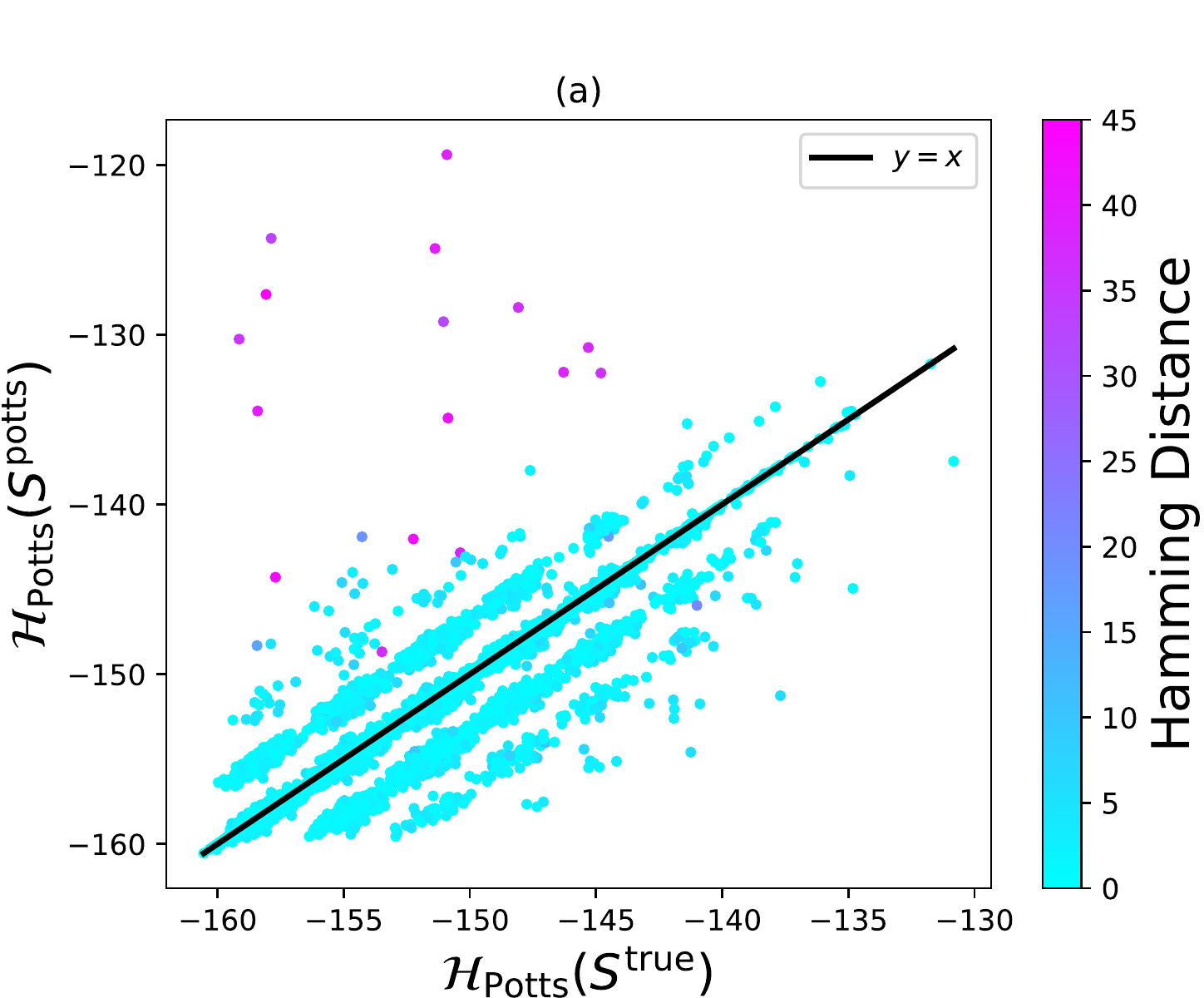}
  \includegraphics[width=.95\columnwidth]{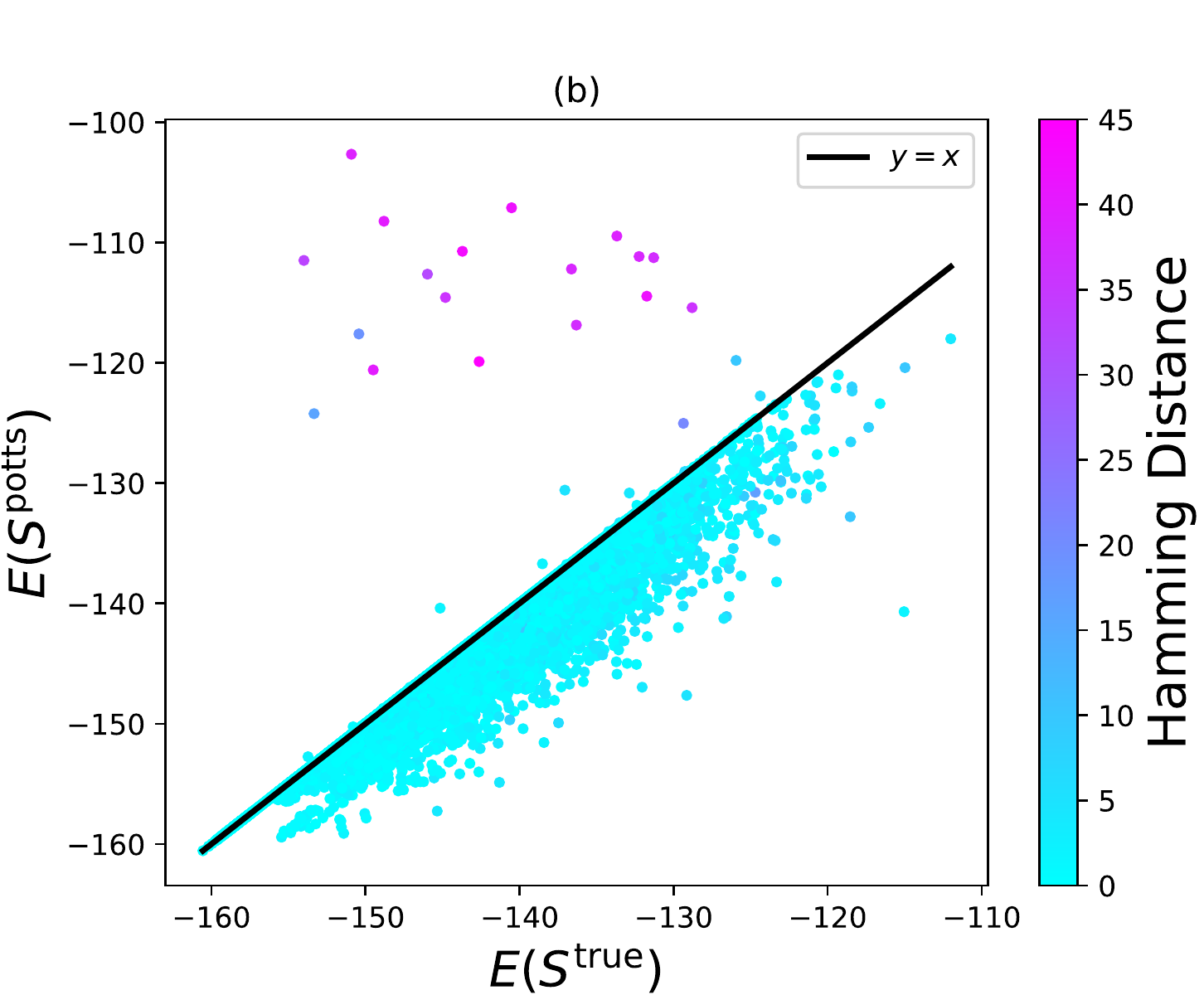}
	\caption{\textbf{Energies of synthetic sequences.} We show
          here the scatter plots of the DCA energy (or Potts
          Hamiltonian) in (a) for the true synthetic sequences
          (x-axis) against the ones aligned by
          DCAlign($\mathcal{H}_{potts}$) (y-axis). In (b), same plot using the
          total cost function $E$.
	\label{fig:en_synth}}
\end{figure}

\subsection{Conservation}
\label{sec:artcons}

The first MSA is generated from a non-trivial profile model, in which
the empirical probability of observing any of the possible amino acids
is position-dependent and it is not uniformly distributed among the
possible states. The generative model used in this case is the profile
model of the PF00018 family, which can be easily learned from the
empirical single site frequencies. From this model we generate
$5\cdot10^{4}$ ``aligned" sequences (the ``ground-truth''), to which
we randomly add some insertions, according to the affine insertion
penalty distributions learned from the PF00018 seed
(Sec.~\ref{sec:MLinspen}), and 20 uniformly randomly chosen symbols at
the beginning and at the end of the aligned sequence.  We split this
alignment into a training set of $2.5\cdot10^{4}$ sequences, which we
use as seed alignment to learn the insertion and gap penalties (using
the scheme for abundant seeds) and a Potts model. We align the
remaining $2.5\cdot10^{4}$ sequences, used as test set.  For
comparison, we build a Hidden Markov Model using $\texttt{hmmbuild}$
of the HMMer package on the training set and we align the test
sequences through the $\texttt{hmmsearch}$ tool.

We show in Fig.~\ref{fig:metrics_synth}(a) the histograms of the
(normalized) Hamming distances, Gap$_+$, Gap$_-$ and Mismatches of the
MSA obtained by DCAlign and HMMer compared to the ground-truth. We
observe that DCAlign is able to find the correct hits and to align
them in a more precise way if compared to HMMer. In fact, the Hamming
distance distribution is shifted to smaller values, suggesting that the
number of errors, per sequence, is smaller than that obtained by
HMMer. The nature of the mistakes seems to be linked to the presence
of mismatches in the case of HMMer, while DCAlign (less often) equally
likely inserts more or less gaps, or a match to the wrong symbol.
While DCAlign is in principle constructed to exploit coevolution,
these results show that even in cases in which, by construction, there
is no coevolution signal, DCAlign is able to perform equally good (or
even better) than state-of-the-art methods.

\subsection{Covariance}

The second experiment is instead focused on correlated data. We ad-hoc
construct an alignment whose first moment statistics resembles that of
a uniform distribution, \ie the probability of observing any
amino-acid, in any column of the seed, is $1/q$. In other words, we
construct the data in such a way that no conserved sites are
present. At the same time, we force the sequences to show coevolving
(\ie correlated) sites, such that the empirical probability of
observing a pair of amino acids is different from that obtained in the
uniform distribution, \ie $f_{ij}(S,S') = \overline{\d_{S_i,S}
  \d_{S_j,S'}} \neq \frac{1}{q^2}$ for some $(i,j)$, where the overline
indicates the empirical average. To construct a dataset with
this statistics, we use as generative model a Potts model with 4
colors (like the RNA alphabet, without the gap state) having non-zero
couplings $J_{ij}(S_i,S_j) = - \d_{S_i,S_j}$ (\ie an
anti-ferromagnetic Potts model) and no fields. The non-zero couplings
are associated with the links of a random regular graph of 50 nodes
and degree 5. The presence of the links ensures the appearance of
non-trivial second moments while, in order to avoid ``polarized''
sites, we sample the model (\ie the Boltzmann distribution associated
with this Hamiltonian) at temperature $T=\frac{1}{\beta} = 0.3$, which
is deep in the paramagnetic phase of this model~\cite{krzkakala2008}.  We perform the same
training pipeline presented in Sec.~\ref{sec:artcons} except for the
learning of the gap penalties: because there are no gap states, we set
$\mu_{\rm ext} = 0$ and $\mu_{\rm{int}} = 0$.

In this case, due to the absence of any conservation,
$\texttt{hmmsearch}$ does not find any eligible hit. In fact, HMMer
tries to align the sequences via a computationally exact recursion on
a HMM, but it has no information to exploit while setting up the HMM
from the training set, because all amino-acids are equally likely to
occur in each column. This represents, of course, the worst-case
scenario for HMM-based methods.  On the contrary, the couplings of the
learned Potts model allow DCAlign to align this kind of sequences. We
remark that contrarily to HMMer, DCAlign has complete information on
the statistics of the training alignment, up to second order
covariances. However, being a heuristic method, it sometimes fails to achieve the (global) minimum of the cost function and
converges to a local minimum, which depends on the initialization
of the target marginals. Re-iterating the MF equations using 10
different seeds of the random number generator suffices to reach the
proper minimum at least once, for the majority of sequences. We remark
that this issue is present only when the MSA does not show any
conserved site and thus the algorithm has no easy ``anchoring" point,
which surely helps lifting the degeneracies in the alignment
procedure. For protein and RNA families presented below the algorithm
seems stable and only one minimum emerges upon re-initialization of
the marginal probabilities of the algorithm.  We quantitatively
measure the performance of DCAlign using four sequence-based metrics
and the energies associated with the aligned sequences. For this
experiment, we refer to the output of our algorithm as the aligned
sequence that has the minimum energy among the 10 trial re-iterations
of the algorithm.  We report the distance metrics in
Fig.~\ref{fig:metrics_synth}(b): the distribution of the Hamming
distances suggests that DCAlign can almost perfectly align the majority of
the target sequences. Indeed, as shown in Fig.~\ref{fig:en_synth}, the
energies (the Potts Hamiltonian alone or the full cost function which
includes the gap and insertion penalties) are identical or very close
to the energies of the true sequences. Only 0.08\% of the aligned
sequences have a Hamming distance density larger than 0.30 (\ie 15
missed positions over 50).
This fraction is so low to be invisible in the histograms of Fig.~\ref{fig:metrics_synth}(b).
 These extreme cases, in which our algorithm
converged to a local minimum in every trial we performed, are
represented as purple points in the scatter plots in
Fig.~\ref{fig:en_synth}.

\begin{table*}[t]
	\centering
	\begin{tabular}{c|c|c|c|c|c|cc|cc|c|c}
		\multirow{2}{*}{Identifier} & \multirow{2}{*}{$L$} & \multirow{2}{*}{$M$ seed} & \multirow{2}{*}{$M_{{\rm eff}}$ seed} & \multirow{2}{*}{pdb} & \multirow{2}{*}{$N_{{\rm seq}}$} & \multicolumn{2}{c|}{$\mu$} & \multicolumn{2}{c|}{$\mu_{{\rm int}}$} & \multirow{2}{*}{mean $N$} & \multicolumn{1}{c}{mean $t$, median $t${[}s{]}}\tabularnewline
		\cline{7-10} \cline{8-10} \cline{9-10} \cline{10-10} \cline{12-12} 
		&  &  &  &  &  & $\mathcal{H}_{\mathrm{potts}}$ & $\mathcal{H}_{\mathrm{phmm}}$ & $\mathcal{H}_{\mathrm{potts}}^{{\rm }}$ & $\mathcal{H}_{\mathrm{phmm}}$ &  & $\mathcal{H}_{\mathrm{potts}}^{{\rm }}$\tabularnewline
		\hline 
		PF00035 & 67 & 81 & 81 & 73 & 19751 & \textcolor{black}{2.50} & \textcolor{black}{2.00} & \textcolor{black}{0.00} & \textcolor{black}{2.00} & 20+$L$+20 & 38, 8\tabularnewline
		\hline 
		PF00677 & 87 & 1878 & 1518 & 9 & 14683 & 0.00 & 0.50 & 2.00 & 2.00 & 20+$L$+20 & 18, 17\tabularnewline
		\hline 
		PF00684 & 67 & 1512 & 1349 & 3 & 10000 & 0.00 & 0.00 & 2.50 & 2.00 & 20+$L$+20 & 20, 7\tabularnewline
		\hline 
		PF00763 & 116 & 1389 & 1355 & 24 & 10000 & 1.50 & 2.50 & 1.00 & 1.50 & 20+$L$+20 & 64, 32\tabularnewline
		\hline 
		RF00162 & 108 & 433 & 241 & 25 & 6026 & 3.50 & 3.50 & 3.00 & 4.00 & 112 & 108, 30\tabularnewline
		\hline 
		RF00167 & 102 & 133 & 105 & 49 & 2631 & 0.50 & 1.50 & 2.00 & 4.00 & 100 & 144, 33\tabularnewline
		\hline 
		RF01734 & 63 & 287 & 287 & 6 & 2017 & 1.00 & 0.00 & 2.00 & 1.50 & 70 & 23, 7\tabularnewline
		\hline 
		RF00059 & 105 & 109 & 83 & 24 & 12558 & 0.00 & 0.50 & 1.50 & 1.50 & 110 & 223, 48\tabularnewline
	\end{tabular}
	\caption{\textbf{Features of the protein and RNA families used
            in this work.} We show here the length $L$ of the
          	sequences for each family, the value of $M$ and $M_{\rm
            eff}$ \cite{morcos2011} for the seed alignment, the number of PDBs used to determine the true contact maps based of real observations of the domains structure, the number of the sequences, $N_{\rm seq}$, to be aligned by our methods and the value of the gap penalties associated with each family and Hamiltonian. For a sub-set of 100 uniformly randomly chosen sequences, we show the average length $N$ of the unaligned sequence (for protein domains, this is set to $20+L+20$); the last column shows the mean and median values of the computing time.
	\label{tab:info_fam} }
\end{table*}

\section{Test on protein and RNA families}

\begin{figure*}[t]
	\centering
	\includegraphics[width=1.0\textwidth]{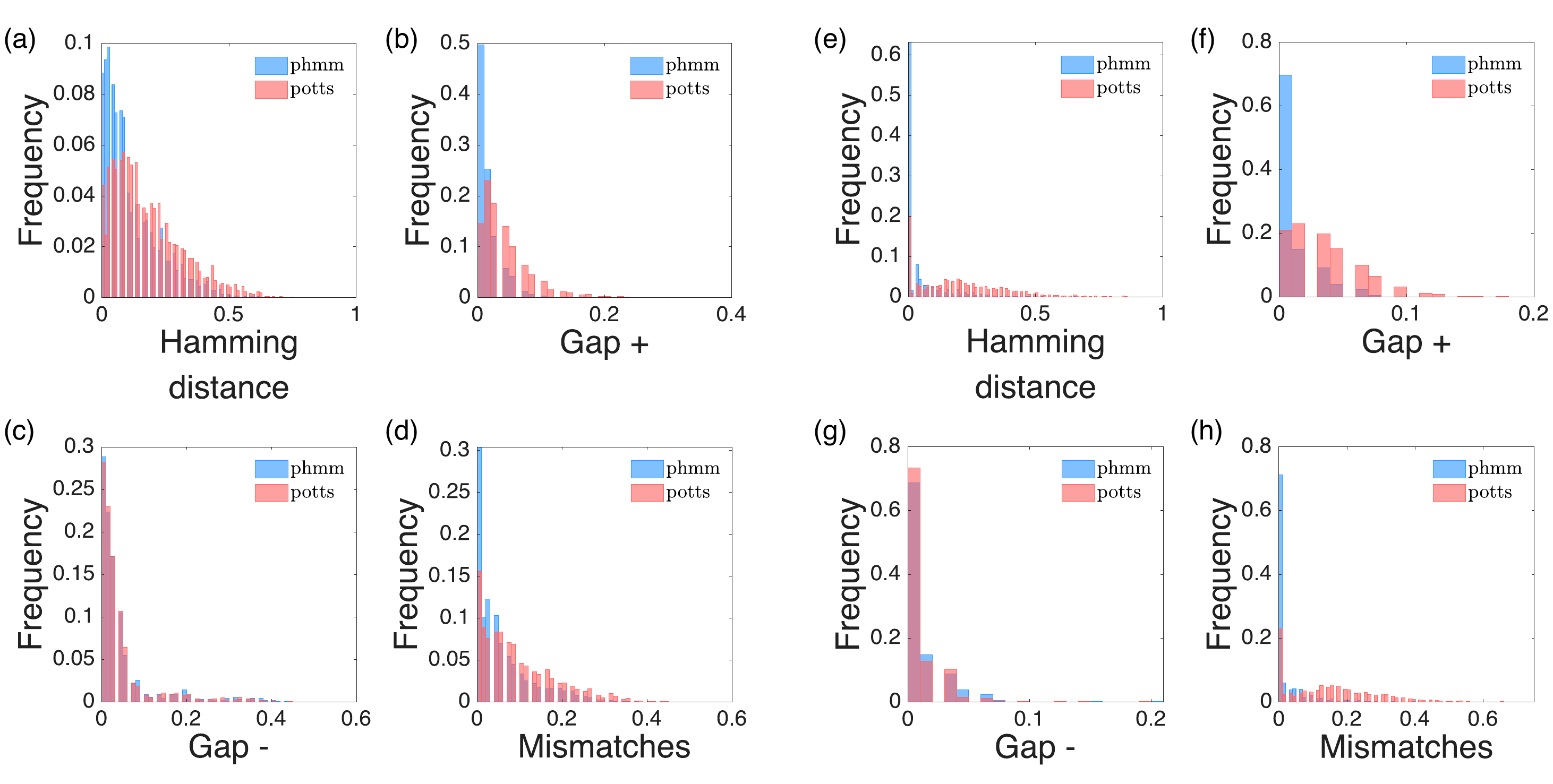}
	\caption{\textbf{DCAlign vs state-of-the-art methods for
            PF00035 and RF00167.}  We plot here the histograms of the
          Hamming distances, Gap$_+$, Gap$_-$ and Mismatches for the
          protein family PF00035 (a, b, c, d) using as reference the
          HMMer results and as target the DCAlign results, and for the
          RNA family RF00167 (e, f, g, h) using as reference the
          Infernal results and as target the DCAlign
          results.  \label{fig:dist_PF35_RF167}}
\end{figure*}

\begin{figure*}[t]
	\centering
	\includegraphics[width=1.0\textwidth]{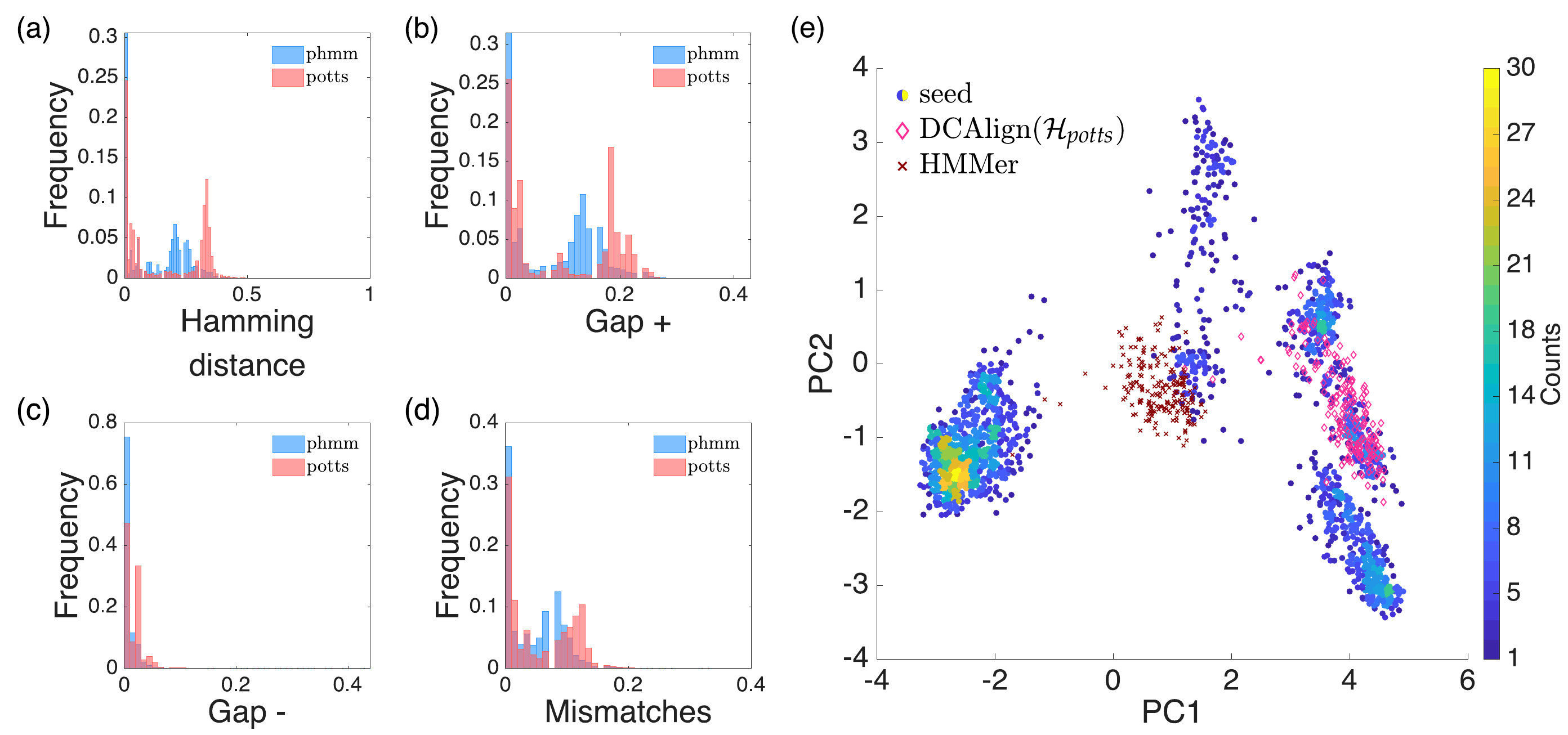}
	\caption{ \textbf{DCAlign vs HMMer for PF00677}. In (a, b, c,
          d) we plot the Hamming distance, Gap$_+$, Gap$_-$ and
          Mismatches using the sequences aligned by HMMer as reference
          and those obtained by DCAlign as target. In (e) we plot
          the projections of the seed sequences in the first two
          principal components of the seed space; the color scale
          denotes the density of the space. The additional sequences
          (depicted as pink diamonds if aligned by
          DCAlign-{\it potts} or as brown crosses if
          aligned by HMMer) are responsible for the red peak around
          0.2 in panel (b).
	\label{fig:dist_677}}
\end{figure*}

\subsection{Choice of families}

We show here the performance of our alignment method for several RNA
and protein families. In particular we select the families PF00035,
PF00677, PF00684, PF00763 from the Pfam database ({\tt
  https:://pfam.xfam.org} release 32.0)~\cite{el2019pfam}, and RF00059,
RF00162, RF00167 and RF01734 from the Rfam database ({\tt
  https:://rfam.xfam.org} release 14.2)~\cite{rfamdb}. The number of
sequences, length of the models, and gap penalties used in the
simulations are reported in Table~\ref{tab:info_fam}.

We restrict our analysis to ``short" families, having $L$ of at most
$100$ positions, in order to avoid a significant slowing down of the
alignment process.  The seed of PF00035 contains very few sequences,
contrarily to PF00677, PF00684, PF00763, which have been chosen
because of their large effective number of seed sequences
$M_{\rm{eff}} > 1000$ (after a standard re-weighting of close-by
sequences \cite{morcos2011}). We thus always infer the gap penalties
according to the abundant seed protocol, except for PF00035. As
reference for comparison, we consider the alignments produced by HMMer
\cite{hmmerweb} and already available in the Pfam database.  We also
perform the alignment of several RNA families, for which
secondary-structure knowledge is necessary to obtain good
alignments with standard tools. We compare our estimate against that obtained by the
state-of-the-art package Infernal \cite{nawrocki2013infernal} which,
indeed, employs the secondary 
structure of the target domains in order to build the so-called
covariance model used to align. Note that, on the contrary,
DCAlign does not use any secondary structure information in the
training procedure (but DCA is able to predict the latter
\cite{de2015direct}). 
As a further comparison we also learn a Hidden
Markov Model (using $\texttt{hmmbuild}$) and we apply
$\texttt{hmmalign}$ to the full length RNA sequences. We choose
precisely these families because of their reasonable length, the
abundance of the seed sequences, and the large number of available
crystal structures, which are useful for the contact map comparison.

\subsection{Comparison with state-of-the-art methods}

As a first comparison, we compute the sequence-based metrics presented
in Sec.~\ref{sec:observables} comparing our full alignment to that
achieved by HMMer, for protein sequences, or by Infernal, when dealing
with RNA families. We show the results for PF00035 in
Fig.~\ref{fig:dist_PF35_RF167}(a-d) and for RF00167 in
Fig.~\ref{fig:dist_PF35_RF167}(e-f), which are representative of the
typical scenario for protein and RNA sequences. The distribution of
all metrics is mostly concentrated in the first bins (the bin width is
set here to 0.01) and decays smoothly at larger distances. The peak in
the first bin is more prominent when the Hamiltonian used for the
alignment is $\textit{phmm}$ indicating that the sequences aligned by
this method are closer to those obtained by HMMer (or Infernal)
than the outcomes of DCAlign-$\textit{potts}$, as
one would expect from the similarity between the two alignment
strategies (see Sec.~\ref{sec:models}).

A notably different behavior emerges for the sequences of the PF00677
family, as shown in Fig. \ref{fig:dist_677}(a-d). It is clear from
Fig.~\ref{fig:dist_677}(a) that a large fraction of the sequences
aligned by DCAlign differs from those aligned by HMMer by about 40\%
of the symbols when using $\mathcal{H}_{potts}$ (the percentage is
reduced to about 30\% when using $\mathcal{H}_{phmm}$). The reason
seems to be partially linked to the presence of mismatches and to a
non-negligible fraction of additional gaps, as indicated by
Gap$_+$. We notice that, differently from the other families, the seed
of PF00677 is composed by several clusters of sequences mostly
differing in the gap composition: a copious fraction of them have
generally few gaps while some other show two long and localized
stretches of gaps. The structure of the seed can be better
characterized in the principal components space, as depicted in
Fig.~\ref{fig:dist_677}(e), where we plot the projections of the seed
sequences in the space of the first two principal components as filled
circles. The colors refer to the density of sequences in the
(discretized) space.  To understand the disagreement between HMMer and
our methods we superimpose the projections of the sequences
responsible for the huge peaks in the Gap$_+$ histogram in the
principal component space of the seed sequences, using brown crosses
for the sequences aligned by HMMer and pink diamonds for those
found by DCAlign-$\textit{potts}$ (note that none
of these sequences is a re-alignment of a seed sequence). Only a small
fraction of the HMMer sequences overlap with the central and poorly
populated cluster while the projections obtained from sequences
aligned by DCAlign-$\textit{potts}$ lie on a well
defined and populated cluster. We thus conclude that looking at the
gap composition of sequences is not sufficient in this case to
understand the different behavior of HMMer and DCAlign. A more
accurate analysis in the principal components space suggests that the
sequences obtained by HMMer are probably miscategorized, at variance
with DCAlign sequences that are in agreement with the seed structure.

In summary, although for some of the families analyzed here (the
distribution of the four metrics for the remaining families are shown
in the SI) the sequences aligned by DCAlign are very similar to those
obtained by HMMer or Infernal, the PF00677 family suggests a different
scenario, in which DCAlign is able to learn some non-trivial correlations present in the seed,
and to exploit them in order to achieve a better alignment of the target sequences.
DCAlign is then able to reproduce state-of-the-art performance in most cases, and to improve them
in some cases.

\subsection{Comparison with the seed}

In this section, we compare the statistical properties of the MSAs
obtained by DCAlign with those of the seed.

\subsubsection{Kullback-Leibler distances}

The statistics of a MSA can be characterized in terms of a statistical
(DCA) model. Depending on the complexity of the model, a certain set
of observables are fitted from the MSA. For instance, in a profile
model only the first moments are fitted, while in a Potts model we can
also fit the information about second moments.  These statistical
models define a probability measure over the space of sequences and
thus characterize a given protein/RNA family.  We consider here the
seed sequences as our ground truth, and we thus consider that a model
learned from the seed is the one that better characterizes the
protein/RNA family under investigation.  We then infer a second model
from the full set of aligned sequences, and we ask how different is
this model from that learned from the seed.  To answer this
question we compute the symmetric Kullback-Leibler divergence $D_{{\rm
    KL}}^{{\rm sym}}$ between the two models, see
Sec.~\ref{sec:observables}, which must be intended as a statistical
measure of distance between the seed and the set of aligned sequences.
In order to fairly compare DCAlign, HMMer and Infernal, which by
construction treat differently the pairwise co-variation of the MSA
sites, we learn, from a seed and from the test alignment, a profile
model $\mathcal{H}^{\rm Prof}$ and a Potts model $\mathcal{H}^{\rm
  Potts}$.

\begin{figure}
	\includegraphics[width=\columnwidth]{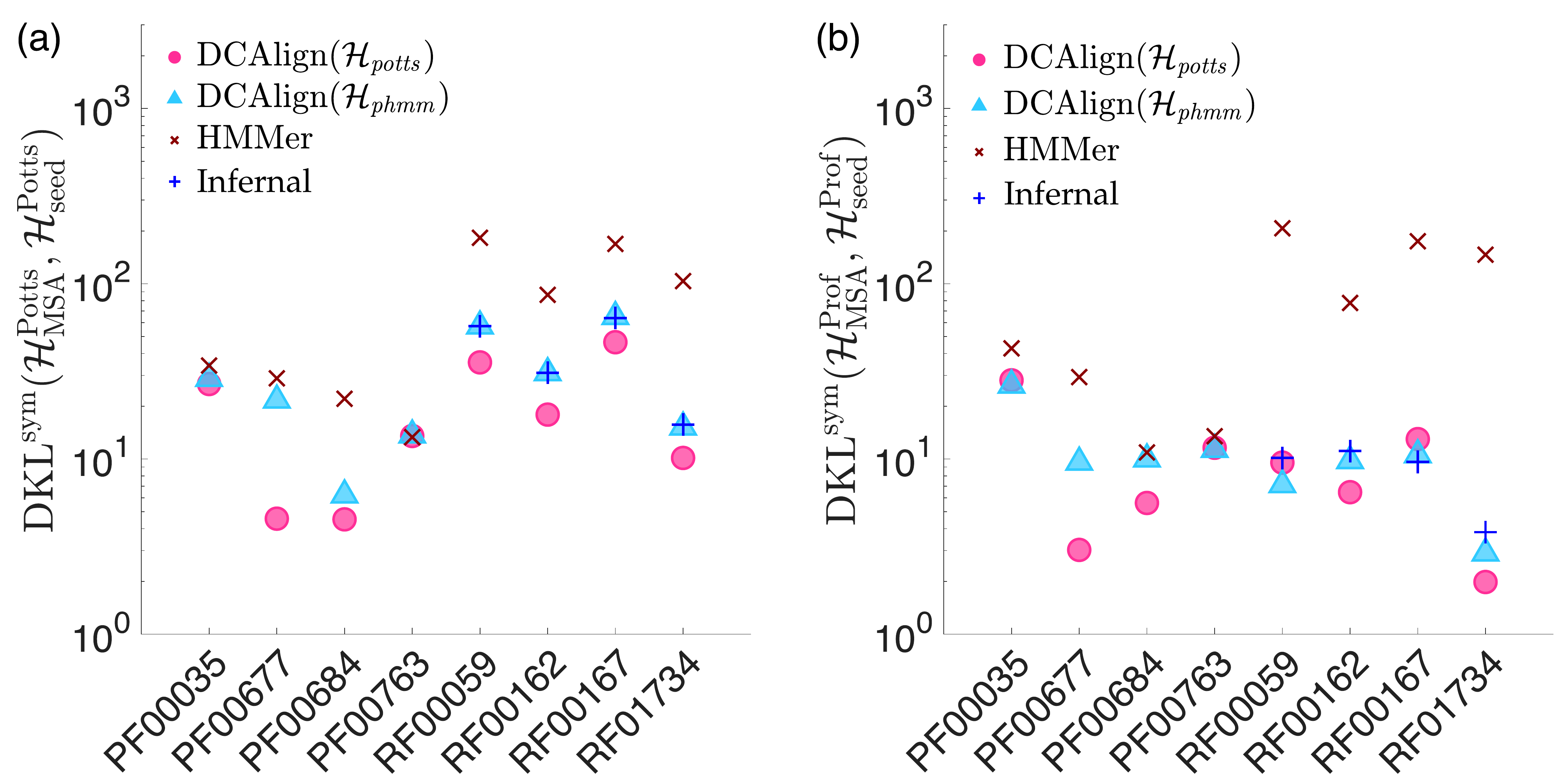}
	\caption{\textbf{Symmetric Kullback-Leibler distances.}  We
          plot the symmetric KL distance between the MSA and the seed
          alignment, computed via a Potts model~(a) or a profile
          model~(b), for all the families and all the alignment
          methods we considered. \label{fig:sDKL}}
\end{figure}

We show in Fig.~\ref{fig:sDKL}(a,b) the results for all families and
all methods, when the model learned is a Potts model or a profile
model, respectively. We notice that the alignments produced by
DCAlign-$\textit{potts}$ always, for the Potts case, and very often,
for the profile case, minimize $D_{{\rm KL}}^{{\rm sym}}$ with respect
to the seed. Infernal is very effective when dealing with RNA
sequences but not as good as DCAlign-$\textit{potts}$ for the majority
of the cases. HMMer always produces the largest distance (except for
PF00763 where basically all methods perform equally good), in
particular for RNA families. We mention that alignments produced by
$\texttt{hmmalign}$ present aligned sequences that always show long
concatenated gaps at the beginning and at the end of the sequence,
differently to the Rfam full alignment, the seed sequences and the
outputs of DCAlign. This partially explains the difference with respect to the
other alignment tools.

These results suggest that the more we use additional information within
the alignment process (in particular, when learning
$\mathcal{H}_\textit{potts}$ we employ all positions and amino-acid
dependent pairwise energy function), the closer the final alignment
will be to the seed. Surprisingly, this feature is retrieved even when
the model learned from the full set of aligned sequences uses less
information than the model used to align, e.g. for the profile model
used in Fig.~\ref{fig:sDKL}(b). Of course, there is a tradeoff, because
including additional statistical properties of the seed in the
alignment process requires a larger seed.

\subsubsection{Proximity measures}
\begin{figure}
	\includegraphics[width=\columnwidth]{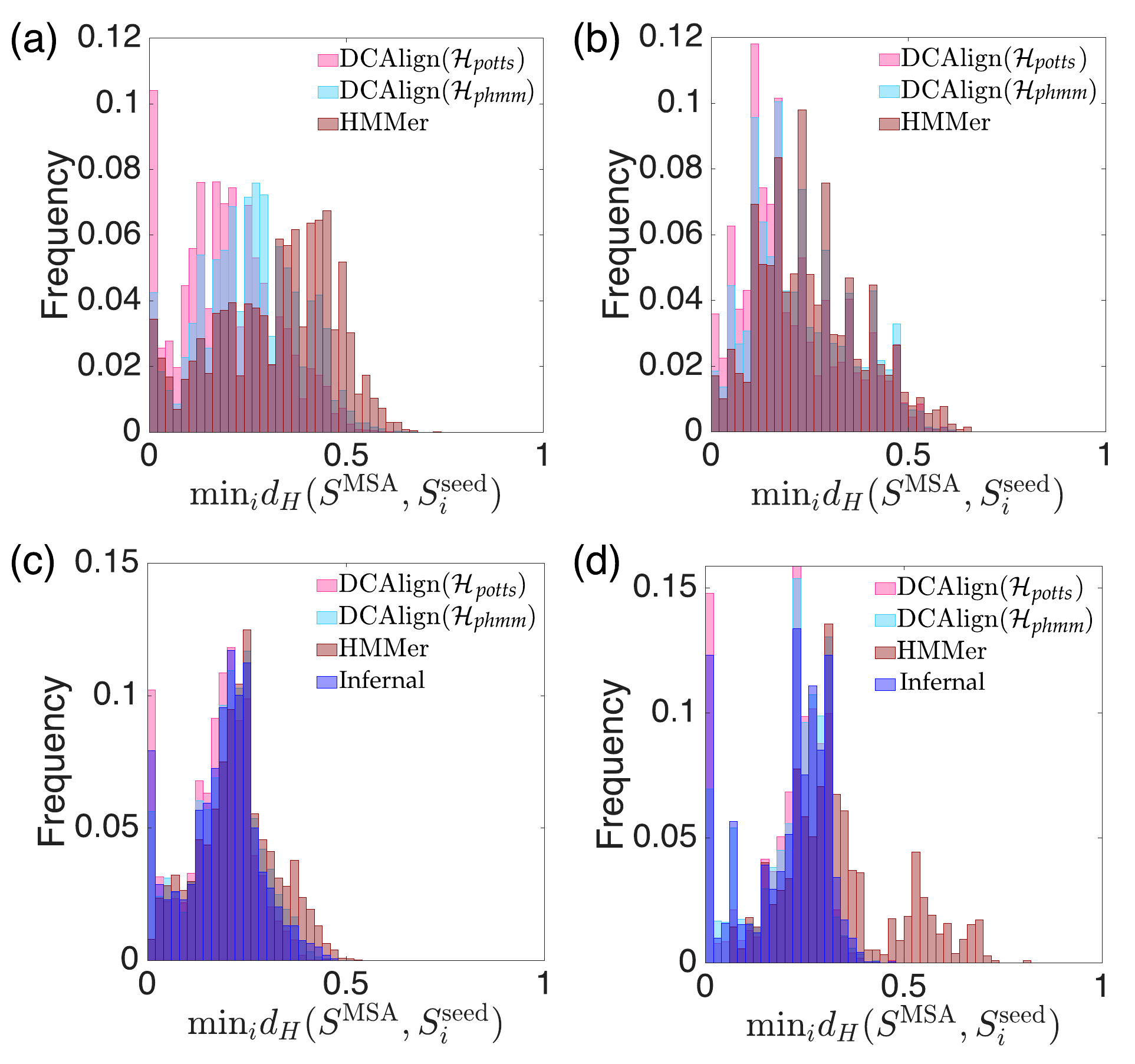}
	\caption{\textbf{Distribution of proximity measures.} Histograms of the minimum distances computed according to Eq.~\eqref{eq:proximity} for the full set of aligned sequences obtained by DCAlign-$potts$, DCAlign-$phmm$, HMMer, and Infernal, against the seed. 
	Panels (a), (b), (c) and (d) refer to the families PF00677, PF00684, RF00162, RF01734 respectively.	\label{fig:proxmeas}}
\end{figure}

We present here a sequence-based comparison between a candidate
alignment, i.e. an alignment obtained by DCAlign-$\textit{potts}$,
DCAlign-$\textit{phmm}$, or HMMer/Infernal, and the seed that will be
considered here as the reference alignment. The metrics we use is the
proximity measure introduced in Sec.~\ref{sec:observables}. We show in
Fig. \ref{fig:proxmeas} the distribution of the minimum distances for
a representative subset of the families, i.e. PF00677 in panel (a),
PF00684 in (b), RF00162 in (c) and RF01734 in (d). Results for
PF00035, PF00762, RF00059 and RF00167 are shown in the SI.  We notice
that for the majority of the families (protein or RNA) the histograms
built from DCAlign-$\textit{potts}$ have a large peak in the first bin
(which collects distances from 0 to 0.02), suggesting that there exist
more sequences in this alignment which are close to the seed than in
any other alignment. A large peak at small distance is also observed
for Infernal when dealing with RNA families, as seen from the blue
histograms in Fig.~\ref{fig:proxmeas}(c) and (d). The Infernal results
overlap quite well with those obtained by DCAlign-$\textit{phmm}$. The
histograms produced by HMMer seem to be shifted to larger Hamming
distances, thus reflecting a smaller similarity to the seed than all
the other methods. Although DCAlign-$phmm$ exploits similar
information to that encoded in HMMer, the corresponding alignment
surprisingly produces, for most of the studied families, results that
are more similar to those obtained by DCAlign-$potts$ or Infernal.

\subsection{Contact prediction}

An important test of the quality of a MSA is related to the
interpretability of the DCA parameters learned from it.  As mentioned in
Sec.~\ref{sec:observables}, the largest couplings are a proxy for the
physical contacts in the folded structure of the protein domains.  In
Table~\ref{table:res_contacts} we report a summary of the results for
three observables associated with the contact prediction: the position
of the first false positive in the ranked Frobenius norms, the value
of the True Positive Rate (TPR) at $2\cdot L$ and the position at
which the TPR is less than 0.80 for the first time. The bold number
corresponds to the largest value, and therefore the best performance,
among all the methods.

\begin{table*} 
	\centering
	\begin{tabular}{c|c|c|c|c|c}
		\multirow{2}{*}{Identifier} & \multicolumn{5}{c}{First FP, TPR(2L), TPR $<$ 0.80}\tabularnewline
		\cline{2-6} \cline{3-6} \cline{4-6} \cline{5-6} \cline{6-6} 
		& $\mathcal{H}^{\rm Potts}_{{\rm seed}}$ & $\mathrm{DCAlign}\left(\mathcal{H_{{\it potts}}}\right)$ & $\mathrm{DCAlign}\left(\mathcal{H}_{{\it phmm}}\right)$ & HMMer & Infernal\tabularnewline
		\hline 
		PF00035 & 9, 0.478, 13 & \textbf{35}, 0.754, 98 & 32, 0.799, \textbf{134} & 28, \textbf{0.791}, 119 & -\tabularnewline
		\hline 
		PF00677 & 22, 0.730, 109 & \textbf{47}, 0.759, 147 & 28, 0.747, 128 & 31, \textbf{0.793}, \textbf{163} & -\tabularnewline
		\hline 
		PF00684 & 20, 0.582, 28 & \textbf{29}, 0.672, 101 & 27, \textbf{0.694}, \textbf{104} & 23, 0.627, 73 & -\tabularnewline
		\hline 
		PF00763 & 80, 0.703, 159 & 89, 0.828, \textbf{288} & 85, 0.836, 254 & \textbf{106}, \textbf{0.849}, 272 & -\tabularnewline
		\hline 
		RF00059 & 18, 0.369, 29 & 29, 0.531, 57 & 29, 0.519, \textbf{64} & \textbf{37}, 0.519, 59 & 33, \textbf{0.566}, \textbf{64}\tabularnewline
		\hline 
		RF00162 & 15, 0.306, 52 & 17, 0.449, \textbf{69} & \textbf{25}, 0.398, 61 & 22, 0.426, 67 & 19, \textbf{0.519}, 61\tabularnewline
		\hline 
		RF00167 & 19, 0.324, 27 & 27, 0.493, 59 & 25, 0.556, 53 & 22, 0.577, \textbf{62} & \textbf{28}, \textbf{0.592}, 57\tabularnewline
		\hline 
		RF01734 & \textbf{10}, 0.300, 16 & \textbf{10}, 0.380, 16 & \textbf{10}, \textbf{0.430}, 16 & \textbf{10}, 0.360, 16 & \textbf{10}, 0.400, \textbf{17}\tabularnewline
	\end{tabular}
	\caption{\textbf{Summary of the contact map results.} For each
          protein or RNA family we show here three metrics computed
          from the PPV curve retrieved from a set of Potts models.
          $\mathcal{H}_{\mathrm{seed}}$ is a Potts model learned using
          the seed sequences alone, while the others are associated
          with the complete alignments obtained by DCAlign-$potts$,
          DCAlign-$phmm$, HMMer and Infernal. The chosen observables
          give the position of the first false positive (First FP),
          the value of the true positive rate (TPR) computed after $2
          \cdot L$ predictions and the rank at which the value of the
          true positive rate is smaller than 0.80 for the first
          time. A perfect prediction is obtained if all the true
          positive contacts are associated with the highest value of
          the Frobenius norm, thus the higher the value of these
          metrics, the better the prediction of the contact maps. We
          show in bold numbers the best performances, for all metrics
          and among all the methods.
		\label{table:res_contacts}}
\end{table*}

\begin{figure*}
	\centering
	\includegraphics[width=1.0\textwidth]{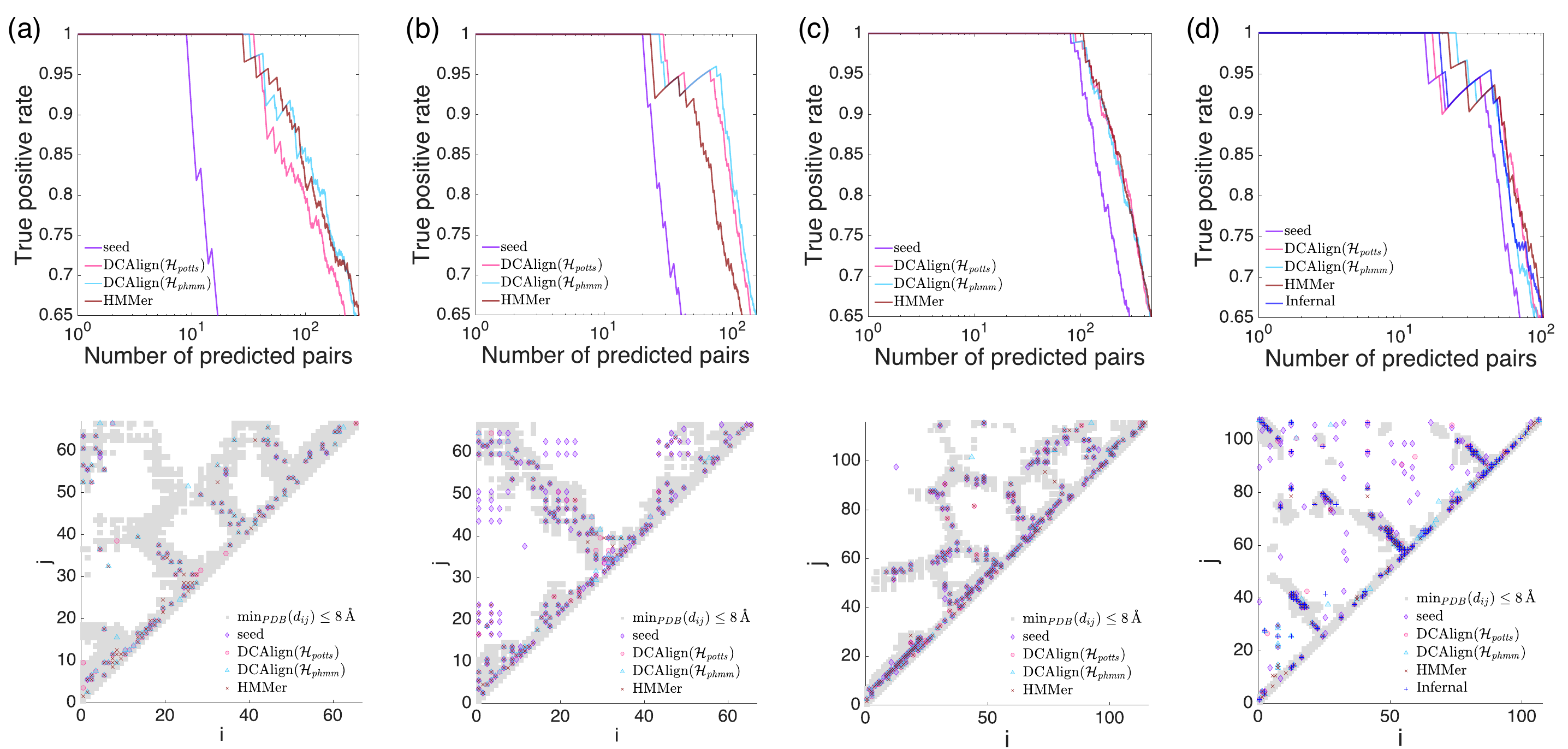}
	\caption{\textbf{Contact predictions.} We show the Positive Predictive Value curves on the top
          panels and, on the bottom ones, the contact map retrieved by
          a set of known crystal structures (gray squares) and the
          Frobenius norms (computed from the full set of aligned
          sequences or the seed), for (a) PF00035, (b) PF00684, (c) PF00763 and
          (d) RF00162. \label{fig:contacts}}
\end{figure*}

We show in Fig.~\ref{fig:contacts} (a,b,c,d) the Positive Predictive
Value (PPV) curves (left) and the contact maps (right) for the
PF00035, PF00684, PF00763 and RF00162 families respectively (results
for PF00677 and the other RNA families are shown in the SI).  The PPV
curves are constructed by plotting the fraction of true positives TP
as a function of the number of predictions (TP and FP),
\ie~PPV=TP/(TP+FP).  The true contact maps are extracted from
all the available PDBs and plotted as gray filled squares, while the
predicted contact maps are constructed by plotting the Frobenius norms
of the DCA couplings that are larger than an arbitrary threshold, here
set to 0.20.  For RNA sequences the comparison between the predictions
and the ground-truth can be performed only using the Frobenius norms
associated with the central part of the aligned sequences, because
there is no available structural information about the sites on the
boundaries.  In addition to the predictions obtained from the full set
of aligned sequences, we show, for comparison, the predicted contact
map obtained from the Potts model inferred from the seed sequences
alone.  As we can notice from Table \ref{table:res_contacts} and the
plot of the contact maps, there is no strategy that clearly
outperforms the others (except the poor results of $\textit{seed}$
which are easily explained by its limited number of sequences). For
RNA families, Infernal seems to accomplish the best predictions in terms of
first FP and TPR but nonetheless all the other methods, included
HMMer, show comparable results. In fact, although HMMer has the
tendency to assign consecutive gaps in the first and last sites of the
aligned sequences, these regions are not considered in the comparison,
and the core part of the alignment suffices to obtain similar results,
in terms of contact prediction, to the other methods.  Although in the
other metrics presented above there was no clear difference between
models learned from large or small seeds, in the contact maps
comparison this seems to be an important issue.  Indeed, the amount of
sequences in the seed slightly affects the quality of the contact map
for our methods: for the PF00035 family (whose seed contains only 81
sequences) DCAlign reaches slightly worse performances than HMMer. On
the contrary, for the PF00763 all methods produce indistinguishable
PPV curves and contact maps.  Finally, we remark the results for
PF00684 in Fig.~\ref{fig:contacts}(b) where DCAlign achieves a better
contact prediction, as manifested by the PPV
lines. This result, not linked to the way of encoding the seed
statistics within the model, but shared by both
$\mathcal{H}_\textit{potts}$ and $\mathcal{H}_\textit{phmm}$, could be
caused by a better treatment of the insertions with respect to HMMer.

\subsection{Running time}

As mentioned in Section \ref{subsec:alignment}, the running time of DCAlign scales roughly quadratically on $L$ and $N$. To give a reference computing time for each family considered here, we aligned $100$ uniformly randomly chosen sequences using a laptop computer, and we measure the mean and the median running time as well as the average length of the full length sequences (note that for protein domains this is fixed to 20 + $L$ + 20 while for RNA domains it varies). We show these quantities in the last columns of Table \ref{tab:info_fam}. Remarkably, the running time is, as expected, affected by the length of the unaligned sequence, but also by the number of sequences of the seed. We notice that the more copious is the seed, the less time (more precisely, the less number of iterations of the MF-based algorithm) is needed to converge. For instance, RF00162 and RF00059 unaligned sequences have roughly the same length $L$ (108 and 102 respectively) and average $N$ (112 and 110) but they differ in the $M_{\rm eff}$ of the seed (241 and 83 respectively); this seems to affect the mean computing time, 108 and 223 seconds for RF00162 and RF00059 respectively.
This suggests that accurate models allow for fast alignment, as DCAlign is able to easily detect the target domains while models learned from ``small'' seeds require more iterations (sometimes the maximum number of iterations, here set to 1000, as in some RNA domains).

\section{Conclusions}

In this work, we have developed and tested DCAlign, a method
to align biological sequences to Potts models of a seed alignment. 
The set of hyper-parameters characterizing the models are inferred by an inverse
statistical-physics based method known as Direct Coupling Analysis,
which captures both the single- and two-site seed
statistics. Single-site statistics often signals residue conservation,
i.e. the propensity of some sites to restrict the variation of residue
(amino-acid or nucleotide) composition because specific residues 
are functionally and/or structurally important at certain
positions.  Two-site statistics is instead related to residue
coevolution: for instance, residues in direct
contact in a folded protein must preserve bio-physically compatible
properties, leading to a correlated evolution of pairs of sites.

Most standard alignment algorithms such as HMMer are based on the assumption of
independent-site evolution, which is statistically encoded via the
so-called profile models, and thus neglect coevolution. In these alignment procedures, strongly
conserved residues serve as anchoring points, and a mismatch in these
positions surely induces bad alignment scores (i.e.~high energies
using a physics-like terminology). Variable sites, characterized by
high entropy values in the seed MSA, do provide little information for
aligning a new sequence to the seed.

However, often residue pairs show a strong degree of coevolution as
reflected in two-site statistics and as a consequence, this important
collective information must be taken into account. Up to now, the only
example in which this information is exploited in the alignment
procedure of RNA sequences, in which 
the base pairing (Watson-Crick or wobble pairs) of the secondary
structure is encoded in the covariance models used to 
align. Note that this structural information must be given as input to
alignment algorithms like Infernal. 

In contrast to more specialized
alignment algorithms like HMMer (using profile HMM) or Infernal (using
covariance models based on secondary RNA structure),
DCAlign takes advantage of both conservation and coevolution
information contained in the seed alignment, and does not require any additional structural input.
The most compatible domain 
among all the possible sub-sequences of a candidate sequence is determined by 
maximizing a score, which 
can be understood as a probability measure of the domain according to
a Boltzmann distribution carefully built from the DCA model and
gap/insertion penalties learned from the seed.  We note that the algorithm
is formulated in a very general way, and it can thus be applied to any kind of sequence,
not necessarily of biological origin.

Using synthetic
data at first, we tested the algorithm under
extreme conditions, when all information is contained in
conservation but none in coevolution, or vice versa. We found that 
DCAlign performs very well in both cases.
This universal applicability is well confirmed in the case of real
data; we tested both protein and RNA sequence data, aligning large
numbers of sequences to the seed MSA provided by the Pfam and Rfam
databases. We find that in most cases, our algorithm performs
comparably well to more specialized state-of-the-art methods, 
while for example profile HMM applied to RNA perform less well. Also, DCAlign does
not need any structural information, being based on the seed statistics only.

Remarkably, in one of the studied
protein families, we find a large group of sequences, which are aligned
differently by HMMer and DCAlign. The sequences aligned
by DCAlign show a better coherence with the seed statistics than those aligned by HMMer, 
as manifested by the Principal Component Analysis in Fig.~\ref{fig:dist_677},
suggesting that
the alignment proposed by DCAlign is to be preferred in this case.

In conclusion, DCAlign provides a general method to solve one of the most important problems
in bioinformatics: aligning individual sequences 
to a reference Multiple Sequence Alignment, taken as a seed.
For the first time, this algorithm includes general coevolution information into the alignment process,
and it can thus be succesfully applied to situations in which conservation is not enough, as we have shown using synthetic data.
Furthermore, DCAlign performs equally well than state-of-the-art methods on protein and RNA data, and outperforms them in some cases. 
The main limitation of DCAlign is that, even if we have shown how to deal with small seeds, a large enough seed alignment is certainly preferable
to obtain more precise coevolution statistics. 

The contact map comparison suggests that the models learned from multiple sequences alignments, aligned by DCAlign, are more accurate in predicting real contacts than the models obtained from the seeds. This is probably due to a more precise description of the features of the target family. Therefore one may think of re-aligning the full set of unaligned sequences using the model learned from the MSA produced by DCAlign, and iterate this procedure to refine the final MSA. This approach would further slow down the alignment process (because it would require to re-determine all the terms of the energy function, including the gap and insertion penalties), nonetheless it could have the advantage of fully exploit the conservation and coevolution signal of the data. We leave this investigation for future work.

On a more technical note, 
we have not discussed here the problem of detecting homologs from a given long sequence, 
but we have restricted our application to the neighborhood of the hits selected by HMMer. 
A possible way of identifying a candidate domain in a long sequence might be running DCAlign for a very few iterations 
on the full sequence and looking at the marginal probabilities of each site: 
the positions associated with the largest probability of matching may correspond to the anchor points of the possible hits. 
We leave this exploration for future work.

The code for aligning to a given seed model is available in \url{https://github.com/infernet-h2020/DCAlign} .

\acknowledgments

The authors thank Sean Eddy, Alessandra Carbone, Francois Coste and
Hugo Talibart for interesting discussions. APM thanks Edoardo Sarti
for interesting discussions and his assistance with the {\it Pfam
  interactions} code. APM, AP, and MW acknowledge funding by the EU
H2020 research and innovation programme MSCA-RISE-2016 under grant
agreement No. 734439 INFERNET.  APM and FZ acknowledge funding from
the Simons Foundation ({\#}454955, Francesco Zamponi) and the access
to the HPC resources of MesoPSL financed by the Region Ile de France
and the project Equip@Meso (reference ANR-10-EQPX-29-01) of the
programme Investissements d’Avenir supervised by the Agence Nationale
pour la Recherche.


%

\clearpage

\input{SI_PRE_v2.tex}

\end{document}

%% file: SI_PRE_v2.tex

\begin{widetext}

\appendix
\section*{Supporting Information}

\subsection{Belief propagation equation}

In order to write the BP equations,
it is convenient to introduce weights $\WW_{i,j}$ associated to pair interactions in the Boltzmann weight Eq.~\eqref{eq:Boltz-weight}.
For $i<j$ we define
\beq
\WW_{i,j}(x_i,n_i,x_{j},n_{j}) = 
\begin{cases}
 e^{J_{i,i+1}(A_{x_{i}n_{i}}, A_{x_{i+1}n_{i+1}})} \chi_{\rm{sr}}(x_i,n_i,x_{i+1},n_{i+1}) \ , & \text{ for } j=i+1 \ , \\
 e^{J_{i,j}(A_{x_{i}n_{i}}, A_{x_{j}n_{j}})} \chi_{\rm{lr}}\left(x_{i},n_{i},x_{j},n_{j}\right) \ , & \text{ for } j > i+1 \ ,
 \end{cases}
\eeq
and for $i>j$ we just set $\WW_{i,j}(x_i,n_{i},x_j,n_{j}) = \WW_{j,i}(x_j,n_{j},x_i,n_{i})$.
The total Boltzmann weight then becomes
\beq
\begin{split}
W(\boldsymbol{x}, \boldsymbol{n}) & = \frac1Z  \prod_{i=1}^L \WW_i(x_i,n_i) \times \prod_{i<j}^{1,L} \WW_{i,j}(x_i,n_i,x_{j},n_{j}) \ ,
\end{split}\eeq
and the BP iteration equations are then written straightforwardly:
\begin{eqnarray}
m_{i \to j}(x_{i}, n_{i}) = \frac{1}{z_{i\to j}} \WW_i(x_i,n_i) \prod_{k\neq\{i,j\}} \sum_{x_{k}, n_{k} }  \WW_{i,k}(x_i,n_i,x_{k},n_{k}) m_{k\rightarrow i}(x_k,n_k) \  ,
\end{eqnarray} 
where $m_{i \rightarrow j}(x_i,n_i)$ are the BP messages. From the converged messages, the marginal probability of node $i$ is estimated as
\begin{eqnarray}\label{eq:SI_P_BP}
P_i(x_{i}, n_{i}) = \frac{1}{z_{i}} \WW_i(x_i,n_i) \prod_{k\neq i} \sum_{x_{k}, n_{k} }  \WW_{i,k}(x_i,n_i,x_{k},n_{k}) m_{k\rightarrow i}(x_k,n_k) \  .
\end{eqnarray} 
Note that if all the couplings vanish for $|i-j|> 1$ and the $\chi_{\rm{lr}}$ are omitted, then $\WW_{ij}=1$ for $|i-j|>1$ and the BP equations reduce to the transfer 
matrix equations, which are exact in that case.
In presence of long range couplings, instead, we stress that the $\chi_{\rm{lr}}$ are redundant and can be omitted in an exact treatment. However, the approximate BP equations
depend on the $\chi_{\rm{lr}}$ and give different results if these terms are omitted. In other words, the BP approximation does not ``commute'' with the insertion/removal
of the $\chi_{\rm{lr}}$ constraints.

\subsection{From belief propagation to mean field}

We can obtain the mean field equations by considering a weak interaction (or large connectivity) limit of the BP equations. 
This limit consists in making two approximations for pairs of sites $j\neq{\{i+1,i-1\}}$ that are not nearest-neighbor in the linear chain:
\begin{enumerate}
\item We assume that $J_{i,j}(A,B)$ is sufficiently small to approximate ${e^{J_{i,j}(A,B)} \simeq 1 + J_{i,j}(A,B)}$. This is likely the case for far away sites that are not in contact in the three-dimensional protein structure. Moreover, when the Hamiltonian is written in zero-sum gauge, all couplings tend to be small 
(often the largest ones are $\sim 1$), because this gauge choice minimizes the Frobenius norm of the coupling matrices $J_{i,j}$. 
\item We approximate the messages $m_{i \rightarrow j} (x_i, n_i)$ with the marginal density $P_{i}(x_i,n_i)$, which is a reasonable choice when $i$ and $j$ and 
far enough that the influence of $j$ on $i$ is negligible.
\end{enumerate}
We emphasize that while both approximations are exact in a mean field setting, in which one takes the thermodynamic limit $L\to\io$ with couplings vanishing with $L$,
they do not hold exactly for our setting in which both $L$ and the couplings are finite.
Moreover, an important remark is that this approximation does not preserve the gauge invariance of the Hamiltonian.

To see the effect of these approximations,
consider, for instance, the marginal density of a node $1< i<L$ in the belief propagation framework. 
Let us call the nearest-neighbor messages $F_i(x_i,n_i) = m_{i\to i+1}(x_i,n_i)$ and $B_i(x_i,n_i) = m_{i\to i-1}(x_i,n_i)$.
The contributions coming from the right and from the left along the linear chains are then
\beq\begin{split}
\FF_i(x_i,n_i) &= \sum_{x_{i-1}, n_{i-1} }  \WW_{i-1,i}(x_{i-1},n_{i-1},x_{i},n_{i}) F_{i-1}(x_{i-1},n_{i-1}) \ , \\
\BB_i(x_i,n_i) &= \sum_{x_{i+1}, n_{i+1} }  \WW_{i,i+1}(x_{i},n_{i},x_{i+1},n_{i+1}) B_{i+1}(x_{i+1},n_{i+1}) \ ,
\end{split}\eeq
and we have
\begin{eqnarray}
P_{i}(x_{i}, n_{i}) = \frac{1}{z_{i}} \mathcal{F}_{i}(x_i,n_i)  \mathcal{B}_{i}(x_i,n_i) \WW_i(x_i,n_i) 
 \prod_{k\neq \{i-1,i,i+1\}} \sum_{\substack{x_{k}, n_{k} : \\ \chi_{\rm{lr}}\left(x_{i},n_{i},x_{k},n_{k}\right) = 1}}  e^{J_{i,k}(A_{x_{i}n_{i}}, A_{x_{k}n_{k}})}  m_{k\rightarrow i}(x_k,n_k) \ .
 \end{eqnarray} 
Applying the weak interaction approximation to the contribution of distant sites,
and introducing
\beq\begin{split}
A_{k\to i}(x_i,n_i) &= \sum_{x_{k}, n_{k} }  \chi_{\rm{lr}}\left(x_{i},n_{i},x_{k},n_{k}\right)  P_{k}(x_k,n_k) \ , \qquad
A_i(x_i,n_i)  = \prod_{k\neq \{i-1,i,i+1\}} A_{k\to i}(x_i,n_i) \ ,
\end{split}\eeq
we obtain
\beq\begin{split}
&P_{i}(x_{i}, n_{i}) \sim \\ &\sim \frac{1}{z_{i}} \mathcal{F}_{i}(x_i,n_i)  \mathcal{B}_{i}(x_i,n_i) \WW_i(x_i,n_i) 
 \prod_{k\neq \{i-1,i,i+1\}} \left( A_{k\to i}(x_i,n_i) + 
 \sum_{\substack{x_{k}, n_{k} : \\ \chi_{\rm{lr}}\left(x_{i},n_{i},x_{k},n_{k}\right) = 1}}  J_{i,k}(A_{x_{i}n_{i}}, A_{x_{k}n_{k}})  P_{k}(x_k,n_k)  \right) \\
 &\sim \frac{1}{z_{i}} A_i(x_i,n_i) \mathcal{F}_{i}(x_i,n_i)  \mathcal{B}_{i}(x_i,n_i) \WW_i(x_i,n_i) 
e^{  \sum_{k\neq \{i-1,i,i+1\}}
\frac{ \sum_{x_k,n_k} \chi_{\rm{lr}}\left(x_{i},n_{i},x_{k},n_{k}\right) J_{i,k}(A_{x_{i}n_{i}}, A_{x_{k}n_{k}})  P_{k}(x_k,n_k) }
{ \sum_{x_k,n_k} \chi_{\rm{lr}}\left(x_{i},n_{i},x_{k},n_{k}\right)  P_{k}(x_k,n_k) }
 } \ .
\end{split}\eeq
Introducing the modified weight
\beq
{\cal C}_{i}(x_{i},n_{i})  =A_i(x_i,n_i)  \WW_i(x_i,n_i) 
e^{\sum_{k\notin\{i,i\pm1\}} \frac{\sum_{x_{k},n_{k}}\chi_{\rm{lr}}(x_{i},n_i,x_k,n_{k})J_{i,k}(A_{x_{i} n_{i}},A_{x_{k} n_{k}})P_{k}(x_{k},n_{k})}
{\sum_{x_{k},n_{k}}\chi_{\rm{lr}}(x_{i},n_i,x_k,n_{k}) P_{k}(x_{k},n_{k})}
 }  \ ,
\eeq
we obtain that 
\beq\label{eq:PFB}
P_{i}(x_{i}, n_{i}) = \frac1{z_i} \mathcal{F}_{i}(x_i,n_i)  \mathcal{B}_{i}(x_i,n_i) \CC_i(x_i,n_i) \ ,
\eeq
which amounts to use the transfer matrix expression
with the replacement $\WW_i \to \CC_i$.
A very similar procedure can be applied to treat the long range part in the forward and backward messages, $F_i$ and $B_i$, with the same result: the equations are 
identical to the transfer matrix equations with $\WW_i \to \CC_i$.
This procedure thus provides a set of closed equations for the forward and backward messages and the marginal probabilities.

For simplicity, we also make an additional approximation, namely that for each pair of distant sites $i,k$ and for each ``relevant'' choice (in a sense that will be more precise below)
of $x_i, n_i$, we have 
\beq\label{eq:approx_norm}
A_{k\to i}(x_i,n_i) = \sum_{x_{k}, n_{k} }  \chi_{\rm{lr}}\left(x_{i},n_{i},x_{k},n_{k}\right)  P_{k}(x_k,n_k) \sim 1 \qquad
\Rightarrow \qquad A_i(x_i,n_i)\sim 1 \ .
\eeq
In words, this amounts to assume that the long-range constraints we artificially introduce do not play a big role on the normalization of marginals, \ie that most of
the mass of $P_k$ is concentrated on compatible assignments of $x_k, n_k$.
Note that for very unlikely values of $x_i, n_i$ (e.g. assigning $x_i=0, n_i=N+1$ at the very beginning of the sequence), the approximation in Eq.~\eqref{eq:approx_norm}
is probably not correct. However, the probability $P_i$ for such values is already suppressed by the short-range terms, making the error irrelevant.
Under this approximation, the modified weights reduce to
\beq
{\cal C}_{i}(x_{i},n_{i})  =  \WW_i(x_i,n_i) 
e^{\sum_{k\notin\{i,i\pm1\}} \sum_{x_{k},n_{k}}\chi_{\rm{lr}}(x_{i},n_i,x_k,n_{k})J_{i,k}(A_{x_{i} n_{i}},A_{x_{k}n_{k}})P_{k}(x_{k},n_{k}) }  \ .
\eeq
The advantage of this additional approximation is that when the long-range couplings $J_{i,j} \to 0$, the mean field equations 
reduce exactly to the transfer matrix equations, as it should be.

\subsection{Mean field free energy}

We give here the expression of the free energy associated with the Boltzmann weight in Eq.~\eqref{eq:Boltz-weight}.
This free energy can be used, for example, as a score 
to compare the quality of the alignment of different sub-regions of the same very long sequence.
The free energy associated with Belief Propagation is given by
\beq
F = - T \sum_i \log z_i + T \sum_{\langle ij\rangle} \log z_{ij} \ ,
\eeq
where the second sum is over distinct pairs $\langle ij \rangle$.
Here, the site term $z_i$ is the denominator in Eq.~\eqref{eq:SI_P_BP}, while the link term is given by
\beq
z_{ij} = \sum_{x_i,n_i,x_j,n_j} m_{i\to j}(x_i,n_i) m_{j\to i}(x_j,n_j)\WW_{i,j}(x_i,n_i,x_j,n_j) \ .
\eeq

In the weak interaction approximation, we can
write the site term $z_i$ as simply the denominator of the single-site marginals, as defined in Eq.~\eqref{eq:PFB},
\ie 
\beq
z_i  = \sum_{x_i, n_i} {\cal C}_{i}(x_i,n_i) {\cal F}_i(x_i,n_i) {\cal B}_i(x_i,n_i) \ .
\eeq
(with small differences on the boundaries).
For the link term $z_{ij}$ we should distinguish between nearest neighboring sites, and far away sites.
For the first case we get
\beq
z_{i-1,i} = \sum_{x_{i-1}, n_{i-1}, x_{i}, n_{i}} \chi_{\rm{sr}}(x_{i-1},n_{i-1},x_{i},n_i) e^{J_{i,i-1}(A_{x_{i-1} n_{i-1}},A_{x_{i} n_{i}})}
F_{i-1}(x_{i-1}, n_{i-1}) B_i(x_i,n_i) \ .
\eeq
For the second case, $j\neq i\pm 1$, we get, 
\beq\begin{split}
	z_{ij} &= \sum_{x_{i}, n_{i}, x_{j}, n_{j}}\chi_{\rm{lr}}(x_{i},n_i,x_{j},n_{j}) e^{J_{ij}(A_{x_{i} n_{i}},A_{x_{j} n_{j}})}
	m_{i\to j}(x_i,n_i) m_{j\to i}(x_j,n_j) \\
	& \simeq \sum_{x_{i}, n_{i}, x_{j}, n_{j}}\chi_{\rm{lr}}(x_{i},n_i,x_{j},n_{j}) e^{J_{ij}(A_{x_{i} n_{i}},A_{x_{j} n_{j}})}
	P_{i}(x_i,n_i) P_{j}(x_j,n_j) \\
	& \simeq \exp\left\{ \sum_{x_{i}, n_i, x_{j}, n_{j}}\chi_{\rm{lr}}(x_{i},n_i,x_{j},n_{j}) J_{ij}(A_{x_{i} n_{i}},A_{x_{j} n_{j}})
	P_{i}(x_i,n_i) P_{j}(x_j,n_j)
	\right\} \ ,
\end{split}\eeq
where we applied the mean field approximation of identifying messages with marginals (first to second line), and considering the interactions $J_{ij}$
as being small (second to third line). Note that in the second step we assumed that
$\sum_{x_{i}, n_{i}, x_{j}, n_{j}}\chi_{\rm{lr}}(x_{i},n_i,x_{j},n_{j}) P_{i}(x_i,n_i) P_{j}(x_j,n_j) = 1$, \ie that the marginals respect the long-range
constraints.

\subsection{Zero temperature limit}

We discuss here briefly how to take the zero temperature limit of the mean field equations.
In order to introduce a temperature $T = 1/\b \neq 1$ we need to rescale all the parameters in the cost function, as $J_{ij} \to \b J_{ij}$, $h_i \to \b h_i$, $\boldsymbol{\mu} \to \b \boldsymbol{\mu}$, $\boldsymbol{\l} \to \b \boldsymbol{\l}$.

In order to take the $T\to 0$ limit we define
\beq\begin{split}
&P_i(x_i,n_i) = e^{\b \pi_i(x_i,n_i)} \ , \qquad F_i(x_i,n_i) = e^{\b \phi_i(x_i,n_i) } \ , \qquad B_i(x_i,n_i) = e^{\b \psi_i(x_i,n_i)} \ , \\
&\CC_i(x_i,n_i) = e^{\b {\rm c}_i(x_i,n_i)} \ , \qquad \FF_i(x_i,n_i) = e^{\b {\rm f}_i(x_i,n_i) } \ , \qquad \BB_i(x_i,n_i) = e^{\b {\rm b}_i(x_i,n_i)} \ .
\end{split}\eeq
We also define $(x_i^*, n_i^*)$ the maximum of $\pi(x_i,n_i)$.
Note that in the first line the messages are normalized, so ${\pi(x^*_i,n^*_i)=0}$, and a similar relation for the other messages. 
In the second line instead, messages are not normalized.
The zero temperature mean field equations are then obtained by taking the limit $\b\to\io$, in which the sums over $x_i, n_i$
 are dominated by the maximum of the integrand. One should only take into account that the hard constraints
$\chi_{\rm{in}}$ and  $\chi_{\rm{end}}$ set to zero some elements of $P_1$, $F_1$, $P_L$ and $B_L$, which translates in $-\io$ elements for $\pi_1$, $\phi_1$, $p_L$ and $\psi_L$.
We obtain (with minor modifications at the boundaries):
\beq\begin{split}
{\rm c}_i(x_i,n_i) & = h_{i}(A_{x_{i}\cdot n_{i}})-\mu(n_{i})(1-x_{i})+\sum_{j\notin\{i,i\pm1\}} \chi_{\rm{lr}}(x_{i},n_{i},x^*_j,n^*_{j})J_{ij}(A_{x_{i}\cdot n_{i}},A_{x^*_{j}\cdot n^*_{j}}) \ , \\
{\rm f}_i(x_i,n_i) & = \max_{ \substack{x_{i-1}, n_{i-1} : \\ \chi_{\rm sr}(x_{i-1},n_{i-1},x_i,n_i) > 0} }
[ \phi_{i-1}(x_{i-1},n_{i-1}) + J_{i-1,i}(A_{x_{i-1}\cdot n_{i-1}},A_{x_i\cdot n_i}) -\f_{i}\left(\Delta n_i\right) \mathbb{I}(n_{i-1}>0) ] \mathbb{I}(n_{i}<N+1) ] \ , \\
{\rm b}_i(x_i,n_i) &= \max_{ \substack {x_{i+1}, n_{i+1} : \\ \chi_{\rm sr}(x_{i},n_{i},x_{i+1},n_{i+1}) > 0} }
[ \psi_{i+1}(x_{i+1},n_{i+1}) + J_{i,i+1}(A_{x_{i}\cdot n_{i}},A_{x_{i+1}\cdot n_{i+1}})-\f_{i+1}\left(\Delta n_{i+1}\right) \mathbb{I}(n_{i}>0) ] \mathbb{I}(n_{i+1}<N+1) ]\ , 
\end{split}\eeq
together with
\beq\begin{split}
\pi_i(x_i,n_i) & ={\rm c}_i(x_i,n_i) +  {\rm f}_i(x_i,n_i) + {\rm b}_i(x_i,n_i) - \max_{x_i,n_i} [{\rm c}_i(x_i,n_i) +  {\rm f}_i(x_i,n_i) + {\rm b}_i(x_i,n_i)]  \ , \\
\phi_i(x_i,n_i) & ={\rm c}_i(x_i,n_i) +  {\rm f}_i(x_i,n_i)  - \max_{x_i,n_i} [{\rm c}_i(x_i,n_i) +  {\rm f}_i(x_i,n_i) ]  \ , \\
\psi_i(x_i,n_i) & ={\rm c}_i(x_i,n_i) +  {\rm b}_i(x_i,n_i)  - \max_{x_i,n_i} [{\rm c}_i(x_i,n_i) +  {\rm b}_i(x_i,n_i) ]  \ .
\end{split}\eeq

\subsection{Maximum likelihood equations for insertions}

We determine the values of  the insertion penalties $\lambda_{o}$, $\lambda_{e}$ by maximizing the likelihood of the data, \ie the $M$ sequences of the seed, given the parameters:
\beq\begin{split}
\mathcal{L}\left(\left\{ \Delta n\right\} _{a=1}^{M}\mid\lambda_{o},\lambda_{e}\right) & = 	\log\prod_{a}P_{i}\left(\Delta n^{a}\mid\lambda_{o},\lambda_{e}\right)-\lambda_{o}^{2}-\lambda_{e}^{2} \\
& = 	\sum_{a}\log P_{i}\left(\Delta n^{a}\mid\lambda_{o},\lambda_{e}\right)-\lambda_{o}^{2}-\lambda_{e}^{2} \\
& = \sum_{a}\left\{ \log\left[\delta_{\Delta n^{a},0}+\left(1-\delta_{\Delta n^{a},0}\right)e^{-\lambda_{o}-\lambda_{e}\left(\Delta n^{a}-1\right)}\right]-\log z\right\} -\lambda_{o}^{2}-\lambda_{e}^{2} \ ,
\end{split}\eeq
where the regularization terms are used to avoid infinite of undetermined parameters.
From the likelihood we obtain an explicit expression of the gradient,
\beq\begin{split}
\frac{\partial\mathcal{L}}{\partial\l_o} &	= 	\sum_{a}\left\{ \frac{-\left(1-\delta_{\Delta n^{a},0}\right)e^{-\l_o-\l_e\left(\Delta n^{a}-1\right)}}{\delta_{\Delta n^{a},0}+\left(1-\delta_{\Delta n^{a},0}\right)e^{-\l_o-\l_e\left(\Delta n^{a}-1\right)}}-\frac{1}{z}\frac{\partial z}{\partial\l_o}\right\} -2\l_o \\
& = 	\sum_{a}\left\{ \frac{-\left(1-\delta_{\Delta n^{a},0}\right)e^{-\l_o-\l_e\left(\Delta n^{a}-1\right)}}{\delta_{\Delta n^{a},0}+\left(1-\delta_{\Delta n^{a},0}\right)e^{-\l_o-\l_e\left(\Delta n^{a}-1\right)}}+\frac{e^{-\l_o}\left(1-e^{-\l_e}\right)^{-1}}{1+e^{-\l_o}\left(1-e^{-\l_e}\right)^{-1}}\right\} -2\l_o \\
& = 	M\left[\frac{e^{-\l_o}\left(1-e^{-\l_e}\right)^{-1}}{1+e^{-\l_o}\left(1-e^{-\l_e}\right)^{-1}}-\left[1-P^{\rm{seed}}(0)\right]-\frac{2}{M}\l_o\right]  \ , \\
\frac{\partial\mathcal{L}}{\partial\l_e} &	= 	\sum_{a}\left\{ \frac{-\left(1-\delta_{\Delta n^{a},0}\right)\left(\Delta n^{a}-1\right)e^{-\l_o-\l_e\left(\Delta n^{a}-1\right)}}{\delta_{\Delta n^{a},0}+\left(1-\delta_{\Delta n^{a},0}\right)e^{-\l_o-\l_e\left(\Delta n^{a}-1\right)}}-\frac{1}{z}\frac{\partial z}{\partial\l_e}\right\} -2\l_e \\
& = 	\sum_{a}\left\{ \frac{-\left(1-\delta_{\Delta n^{a},0}\right)\left(\Delta n^{a}-1\right)e^{-\l_o-\l_e\left(\Delta n^{a}-1\right)}}{\delta_{\Delta n^{a},0}+\left(1-\delta_{\Delta n^{a},0}\right)e^{-\l_o-\l_e\left(\Delta n^{a}-1\right)}}+\frac{e^{-\l_o-\l_e}\left(1-e^{-\l_e}\right)^{-2}}{1+e^{-\l_o}\left(1-e^{-\l_e}\right)^{-1}}\right\} -2\l_e \\
& = 	M\left[\frac{e^{-\l_o-\l_e}\left(1-e^{-\l_e}\right)^{-2}}{1+e^{-\l_o}\left(1-e^{-\l_e}\right)^{-1}}-\left\langle \Delta n\right\rangle _{P^{{\rm seed}}\left(\Delta n\right)}+\left[1-P^{\rm{seed}}(0)\right]-\frac{2}{M}\l_e\right] \ .
\end{split}\eeq
The maximum likelihood estimators can then be obtained by gradient descent, as detailed in the main text.

\clearpage

\subsection{Sequence-based distances plot}

\begin{figure*}[b]
	\centering
	\includegraphics[height=0.8\textheight]{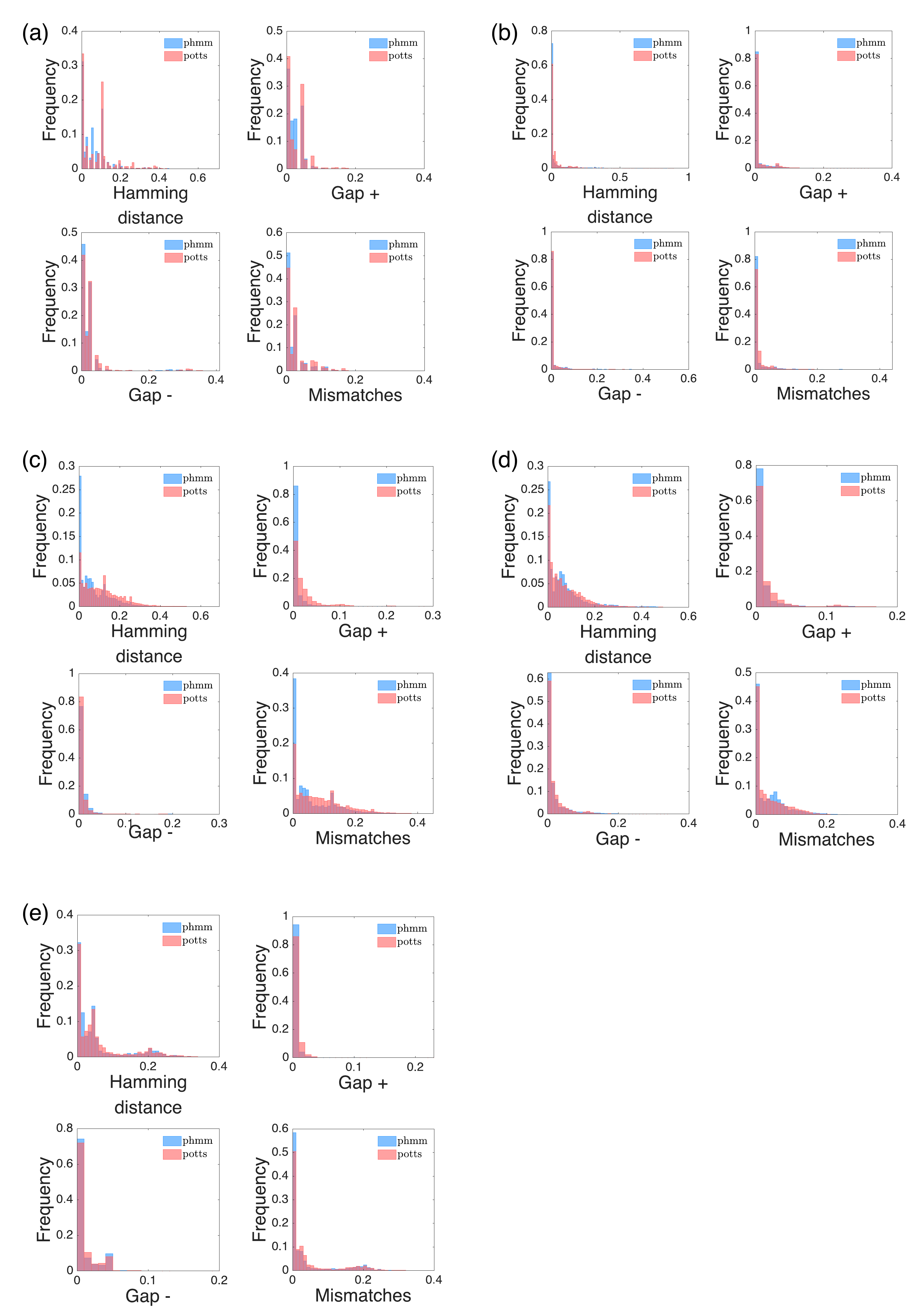}
	\caption{\textbf{Distances distribution.} We plot here the histograms of the
		Hamming distances, Gap$_+$, Gap$_-$ and Mismatches for the
		protein families PF00684 (a), PF00763 (b) using as reference the
		HMMer results and as target the DCAlign results, and for RF00059 (c), RF00167 (d) and RF01734 (e) using as reference the
		Infernal results and as target the DCAlign
		results. 	\label{fig:SI_dist}}
\end{figure*}

\clearpage

\subsection{Proximity measures plot}

\begin{figure*}[h]
	\centering
	\includegraphics[width=0.7\textwidth]{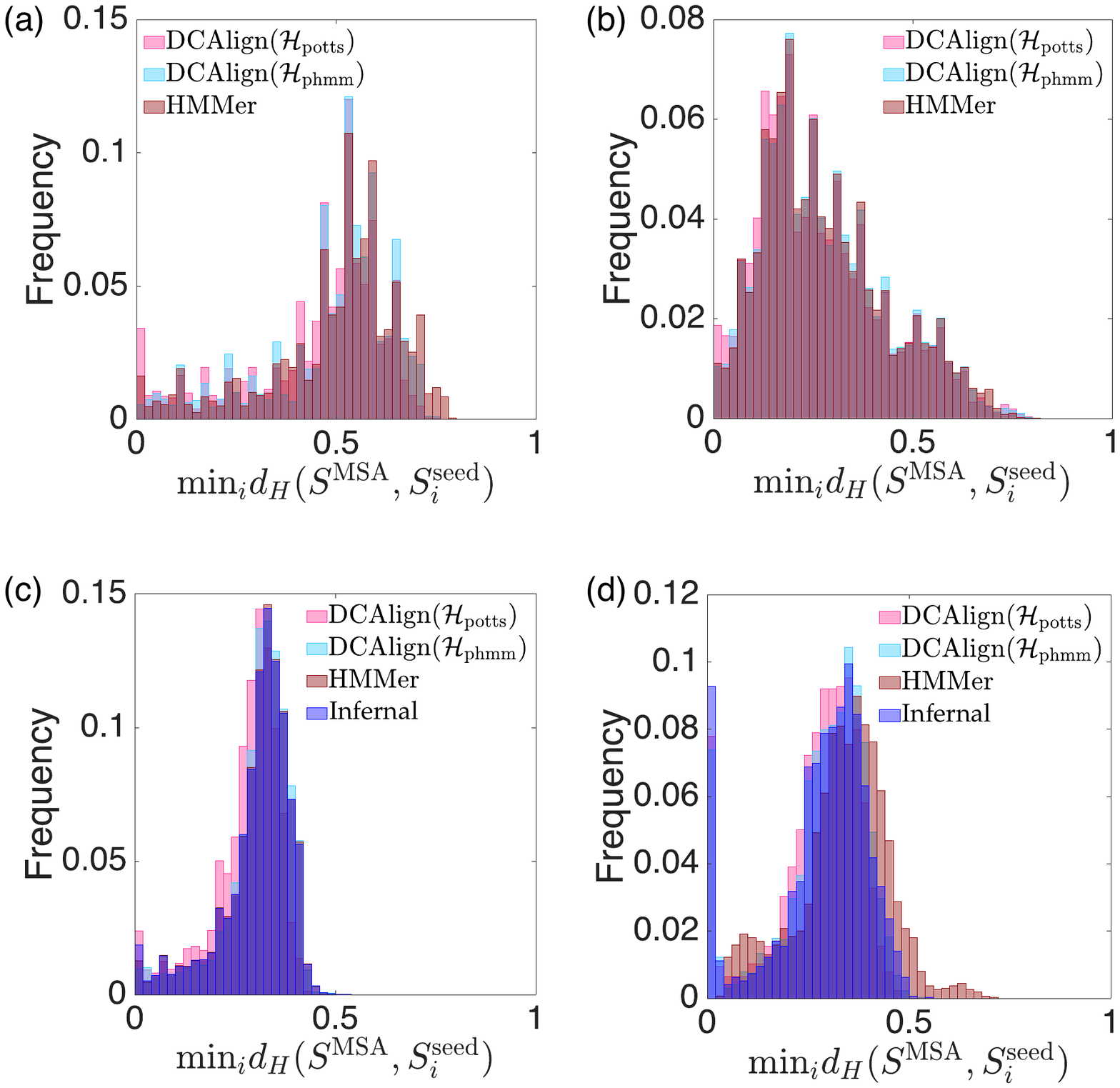}
	\caption{\textbf{Proximity measures.} Histograms of the minimum distances computed according to Eq.~\eqref{eq:proximity} for the full set of aligned sequences obtained by DCAlign-$potts$, DCAlign-$phmm$, HMMer, and Infernal, against the seed. 
	Panels (a), (b), (c) and (d) refer to the families PF00035, PF00763, RF00059, RF00167 respectively. \label{fig:SI_prox}}
\end{figure*}

\clearpage

\subsection{Contact maps}

\begin{figure*}[h]
	\centering
	\includegraphics[width=1.0\columnwidth]{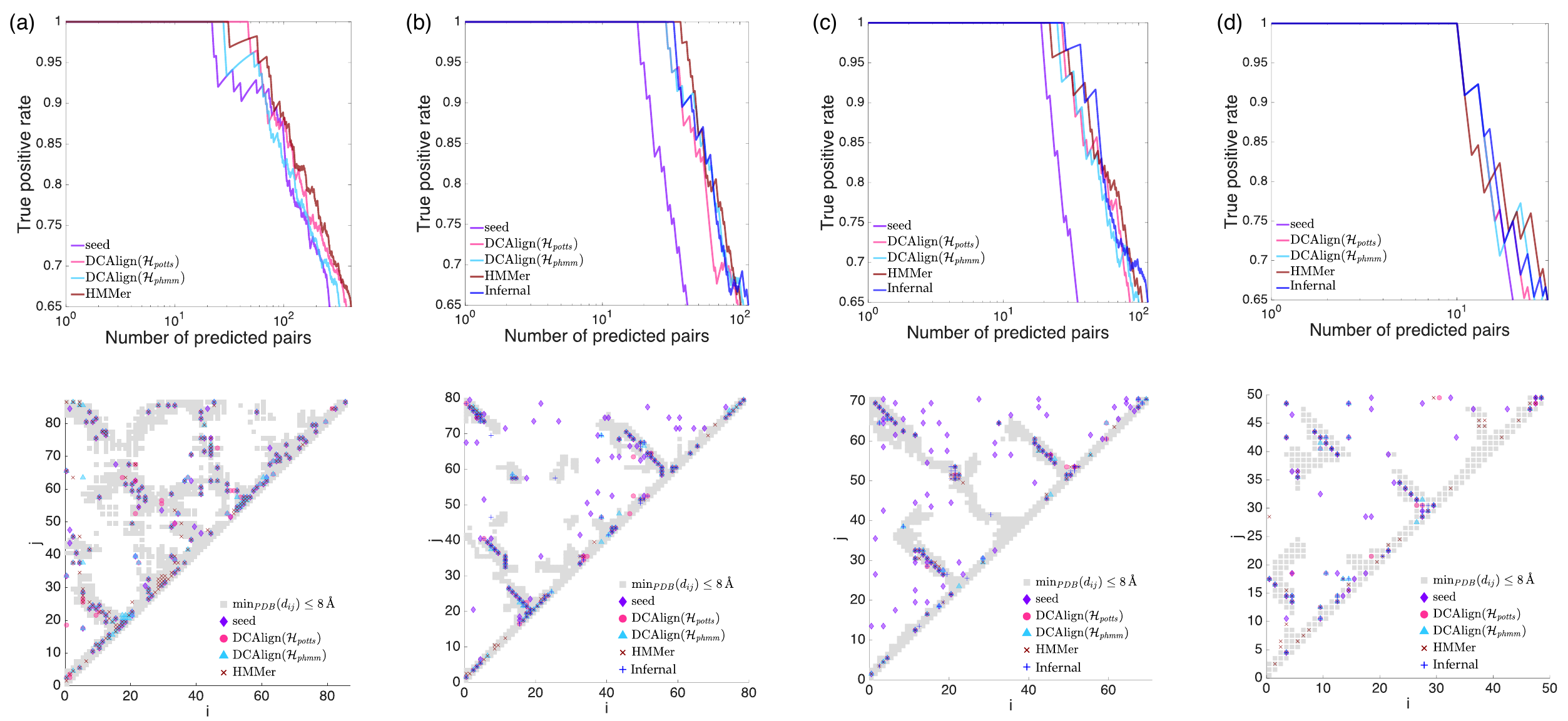}
	\caption{\textbf{Contact predictions.} We show the Positive Predictive Value curves on the top
          panels and, on the bottom ones, the contact map retrieved by
          a set of known crystal structures (gray squares) and the
          Frobenius norms (computed from the full set of aligned
          sequences or the seed), for (a) PF00677, (b) RF00059, (c) RF00167 and
          (d) RF01734. \label{fig:SI_contacts}}
\end{figure*}

\end{widetext}